\documentclass[cernpreprint, atlasdraft=false, texlive=2023, UKenglish, texmf, orcidlogo]{atlasdoc}
 
\usepackage{atlaspackage}
\usepackage{multirow}
\usepackage{amsmath}
 
\usepackage[backref=false]{atlasbiblatex}
 
\usepackage[BSM=true,jetetmiss=true,process=true]{atlasphysics}

\graphicspath{{logos/}{figures/}}
 
\usepackage{ANA-EXOT-2022-15-PAPER-defs}

\AtlasTitle{Search for light long-lived neutral particles from Higgs boson decays via vector-boson-fusion production from $pp$ collisions at $\sqrt{s}=13$ TeV with the ATLAS detector}

\AtlasAbstract{
A search is reported for long-lived dark photons with masses between 0.1 GeV and 15 GeV,
from exotic decays of Higgs bosons produced via vector-boson-fusion.
Events that contain displaced collimated Standard Model fermions
reconstructed in the calorimeter or muon spectrometer are probed.
This search uses the full LHC Run 2 (2015--2018) data sample collected in proton--proton collisions at $\sqrt{s}=13$ TeV,
corresponding to an integrated luminosity of 139 $fb^{-1}$.
Dominant backgrounds from Standard Model processes and non-collision sources are estimated using data-driven techniques.
The observed event yields in the signal regions are consistent with the expected background.
Upper limits on the Higgs boson to dark photon branching fraction are reported
as a function of the dark photon mean proper decay length
or of the dark photon mass and the coupling between the Standard Model and the potential dark sector.
This search is combined with previous ATLAS searches obtained in the gluon--gluon fusion and \textit{WH} production modes.
A branching fraction above 10\% is excluded at 95\% CL for a 125 GeV Higgs boson decaying into two dark photons for dark photon
mean proper decay lengths between 173 and 1296 mm and mass of 10 GeV.
}

\PreprintIdNumber{CERN-EP-2023-226}

\AtlasJournal{EPJC}
\AtlasJournalRef{Eur. Phys. J. C 84 (2024) 719}
\AtlasDOI{DOI:110.1140/epjc/s10052-024-12902-7}

\hypersetup{pdftitle={ATLAS document},pdfauthor={The ATLAS Collaboration}}
 
\begin{document}
 
\maketitle

\section{Introduction}
\label{sec:intro}
 
Many theories beyond the Standard Model (BSM) predict the existence of dark sectors that are weakly coupled to the Standard Model (SM)~\cite{ArkaniHamed:2008qp,Baumgart:2009tn,Falkowski:2014ffa,Curtin:2014cca}.
Depending on how the dark sector is structured and how it couples to the SM, unstable dark states could potentially be generated at colliders, which could then decay into SM particles with significant branching fractions.
 
The scenario explored in this paper introduces a dark photon (\darky), acting as
the mediator of a broken dark U(1) gauge interaction in the dark sector, and mixing kinetically with the SM hypercharge.
These dark photons can be produced in the exotic decay of the Higgs boson and subsequently decay into pairs of
leptons and light quarks~\cite{Cheung:2009su,Falkowski:2010cm,Falkowski:2010gv}.
The latest LHC measurements on the Higgs boson properties do not rule out decays to yet undetected states with
a potentially sizable branching ratio of up to 12\%~\cite{HIGG-2021-23}.
The mean proper lifetime ($\tau$) of the \darky is inversely proportional to the mass of the dark photon and to the square of the kinetic mixing parameter ($\epsilon$)~\cite{Cheung:2009qd, Batell:2009yf}, which is theoretically allowed to vary over values $\epsilon \lesssim 10^{-2}$.
This study focuses on the dark photons with masses in the $\mathcal{O}$(\mev--\gev) range and small values of $\epsilon$ ($<10^{-5}$),
where  decays of the \darky can occur at a macroscopic distance from  their production point.
 
The decay branching fractions of a light dark photon with kinetic mixing with the SM photon
depend on its mass~\cite{Falkowski:2010cm, Meade:2009rb, Batell:2009yf}.
Dark-photon masses smaller than twice the electron mass result in an invisible Higgs signature and are not considered in this paper.
In the range between twice the electron mass and twice the muon mass, the dark photon decays exclusively into electrons, while below twice the pion mass, it decays with equal probability into pairs of muons and electrons.
In the range above twice the pion mass, dark photons can decay to electrons, muons, and hadrons,
where decay branching fractions vary as a function of the mass due to the presence of hadronic resonances~\cite{Zyla:2020zbs}.
Due to the large Lorentz boosts expected for dark photons produced with small masses relative to the energy scale of the hard-scattering process, their decay products are expected to be a collimated group of fermions forming a  structure similar to a jet, which is referred to as dark-photon jets (\LJs).
 
\begin{figure}[ht!]
\centering
\includegraphics[width=.42\textwidth]{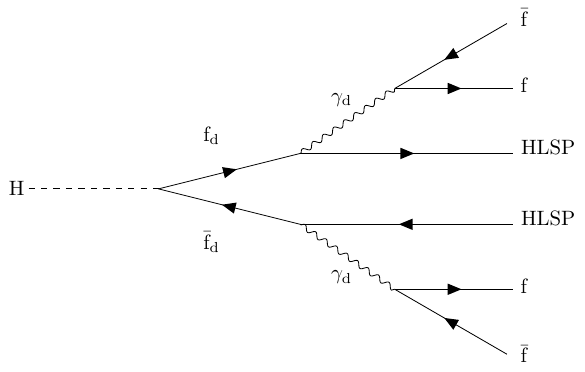}\label{fig:models:a}
\caption{FRVZ process in Higgs boson decays, with the dark fermion \fd decaying into a \darky and an HLSP.
The \darky\ decays into SM fermions, denoted by $f$ and $\bar{f}$. }
\label{fig:models}
\end{figure}

The Falkowski-Ruderman-Volansky-Zupan (FRVZ) model~\cite{Falkowski:2010cm, Falkowski:2010gv} is used to optimise event selection and interpret the results.
In this model, a pair of dark fermions \fd is produced through a Higgs boson decay and each
decays promptly into a dark photon and a stable dark fermion, assumed to be the undetected hidden lightest stable particle (HLSP).
This leads to final states with two dark photons as shown in Figure~\ref{fig:models}.
The Higgs boson is assumed to be SM-like with a mass of 125 GeV.
The Hidden Abelian Higgs Model (HAHM)~\cite{Curtin:2014cca}, which predicts a direct decay of the Higgs boson into a pair of \darky, and commonly used as an alternative model to the FRVZ is not considered. This omission is due to the lack of intrinsic missing transverse momentum in such signal events, which limits the analysis sensitivity.
Dark-sector radiation~\cite{Strassler:2006im} can produce additional dark photons in proportion to the size of the dark gauge coupling $\alpha_\textrm{d}$~\cite{Cheung:2009su}.
To limit the amount of additional \darky emissions, a  dark coupling $\alpha_\textrm{d} \lesssim 0.01$ is assumed in this paper.
Despite not making explicit use of this signal feature,
the analysis remains sensitive to new physics scenarios that involve multiple \darky emissions due to the inclusive number of constituents within the \LJ. However, specific interpretation for these scenarios are not provided.
 
The search for displaced \LJs presented uses the data sample collected at the Large Hadron Collider (LHC) by the ATLAS detector during 2015--2018 in proton--proton ($pp$) collisions at $\sqrt{s}=13$~\tev, corresponding to an integrated luminosity of 139~\ifb.
Previous ATLAS searches for displaced \LJs were performed using $pp$ collision data
at centre-of-mass energies of both 8~\tev~\cite{EXOT-2011-12, EXOT-2013-22} and 13~\tev~\cite{EXOT-2017-28,EXOT-2019-05},
exploring the gluon--gluon fusion (ggF) and \textit{WH} production modes.
These searches provide complementary results to those from related ATLAS
searches for prompt \LJs~\cite{EXOT-2011-01,EXOT-2012-16,EXOT-2014-09} probing higher values of $\epsilon$,
and for displaced dimuon vertices~\cite{EXOT-2017-03} targeting larger \darky mass values.
Related searches for dark photons were conducted by the CDF and D0 collaborations at the Tevatron~\cite{Aaltonen:2012pu,Abazov:2009hn,Abazov:2010uc} and by the CMS~\cite{CMS-EXO-11-013,CMS-EXO-12-012,CMS-HIG-13-010,CMS-EXO-19-007}, and LHCb~\cite{Aaij:2015tna,LHCb:2019vmc} collaborations at the LHC\@.
A wide range of constraints on dark photon productions can also be extracted
from beam-dump and fixed-target experiments~\cite{Blumlein:2011mv,Bjorken:2009mm,Bross:1989mp,Merkel:2011ze,Adlarson:2013eza,Abrahamyan:2011gv,Reece:2009un,Blumlein:2013cua,Gninenko:2012eq,Essig:2010gu,Agakishiev:2013fwl}, $e^{+}e^{-}$ collider experiments~\cite{Archilli:2011zc,Babusci:2012cr,Aubert:2009cp,Lees:2014xha,Lees:2015rxq,TheBelle:2015mwa,Won:2016pjz,Ablikim:2018bhf,Ablikim:2017aab}, electron and muon anomalous magnetic moment measurements~\cite{Pospelov:2008zw,Davoudiasl:2012ig,Endo:2012hp}, and astrophysical observations~\cite{Chang:2018rso,Dreiner:2013mua}.
 
Building on the Run 2 ATLAS search for displaced \LJs~\cite{EXOT-2019-05},
this paper reports for the first time studies using the vector-boson-fusion (VBF) Higgs-boson production mechanism.
This leads to a distinctive topology, characterised by
a pair of highly energetic quark-induced jets separated by a significant pseudorapidity gap, resulting in a large invariant mass.
Due to the reduction of SM backgrounds in the VBF channel, signal regions requiring as few as one \LJ are feasible,
extending the sensitivity to dark photons with shorter and longer decay lengths ($c\tau$).
A statistical combination with the ggF and \textit{WH} production channels~\cite{EXOT-2019-05}
is performed to maximise the search sensitivity to the FRVZ model.


\FloatBarrier

\section{ATLAS detector}
\label{sec:detector}
\newcommand{\AtlasCoordFootnote}{
ATLAS uses a right-handed coordinate system with its origin at the nominal interaction point (IP) in the centre of the detector and the \(z\)-axis along the beam pipe. The \(x\)-axis points from the IP to the centre of the LHC ring, and the \(y\)-axis points upwards. Cylindrical coordinates \((r,\phi)\) are used in the transverse plane, \(\phi\) being the azimuthal angle around the \(z\)-axis. The pseudorapidity is defined in terms of the polar angle \(\theta\) as \(\eta = -\ln \tan(\theta/2)\), and the rapidity is defined as \(y = (1/2) [(E+p_z)/(E-p_z)]\).}
 
The ATLAS detector~\cite{PERF-2007-01} at the LHC covers nearly the entire solid angle around the collision point.\footnote{\AtlasCoordFootnote}
It consists of an inner tracking detector surrounded by a thin superconducting solenoid, electromagnetic and hadron calorimeters,
and a muon spectrometer incorporating three large superconducting air-core toroidal magnets.
 
The inner-detector system (ID) is immersed in a \qty{2}{\tesla} axial magnetic field
and provides charged-particle tracking in the range \(|\eta| < 2.5\).
The high-granularity silicon pixel detector covers the vertex region and typically provides four measurements per track,
the first hit normally being in the insertable B-layer (IBL) installed before Run~2~\cite{ATLAS-TDR-19,PIX-2018-001}.
It is followed by the silicon microstrip tracker (SCT), which usually provides eight measurements per track.
These silicon detectors are complemented by the transition radiation tracker (TRT),
which enables radially extended track reconstruction up to \(|\eta| = 2.0\).
The TRT also provides electron identification information
based on the fraction of hits (typically 30 in total) above a higher energy-deposit threshold corresponding to transition radiation.
 
The calorimeter system covers the pseudorapidity range \(|\eta| < 4.9\).
Within the region \(|\eta|< 3.2\), electromagnetic calorimetry is provided by barrel and
endcap high-granularity lead/liquid-argon (LAr) calorimeters,
with an additional thin LAr presampler covering \(|\eta| < 1.8\)
to correct for energy loss in material upstream of the calorimeters.
Hadron calorimetry is provided by the steel/scintillator-tile calorimeter,
segmented into three barrel structures within \(|\eta| < 1.7\), and two copper/LAr hadron endcap calorimeters.
The solid angle coverage is completed with forward copper/LAr and tungsten/LAr calorimeter modules
optimised for electromagnetic and hadronic energy measurements respectively.
 
The muon spectrometer (MS) comprises separate trigger and
high-precision tracking chambers measuring the deflection of muons in a magnetic field generated by the superconducting air-core toroidal magnets.
The field integral of the toroids ranges between \num{2.0} and \qty{6.0}{\tesla\metre}
across most of the detector.
Three layers of precision chambers, each consisting of layers of monitored drift tubes, cover the region \(|\eta| < 2.7\),
complemented by cathode-strip chambers in the forward region, where the background is highest.
The muon trigger system covers the range \(|\eta| < 2.4\) with resistive-plate chambers in the barrel, and thin-gap chambers in the endcap regions.
 
Interesting events are selected by the first-level trigger system implemented in custom hardware,
followed by selections made by algorithms implemented in software in the high-level trigger~\cite{TRIG-2016-01}.
The first-level trigger accepts events from the \qty{40}{\MHz} bunch crossings at a rate below \qty{100}{\kHz},
which the high-level trigger further reduces in order to record events to disk at about \qty{1}{\kHz}.
 
An extensive software suite~\cite{ATL-SOFT-PUB-2021-001} is used in data simulation, in the reconstruction
and analysis of real and simulated data, in detector operations, and in the trigger and data acquisition
systems of the experiment.


\FloatBarrier

\section{Data and simulated event samples}
\label{sec:samples}
Data were collected using the ATLAS detector during Run~2 of the LHC from 2015 to 2018, at a peak instantaneous luminosity of $2.1\times 10^{34}$cm$^{-2}$s$^{-1}$. This resulted in an average of $\langle \mu \rangle = 34$ $pp$ interactions per bunch crossing (pile-up). To ensure  the
normal operation of all the subdetectors and a stable-collision mode for the LHC beams, data quality requirements were applied. The resulting data sample has an integrated luminosity of $139.0 \pm 2.4$~\ifb~\cite{DAPR-2018-01}.
 
In the \mLJ channel, events were selected using two types of MS-based triggers or unprescaled missing transverse momentum (\MET)
triggers~\cite{TRIG-2019-01}, to target the long-lived \darky decays into muons or signature of the undetected HLSP.
The two types of MS-based triggers,  `tri-muon MS-only' and `muon narrow-scan'~\cite{TRIG-2012-02}
were the same as those developed in the ggF production mode~\cite{EXOT-2019-05}.
The narrow-scan trigger efficiencies for signals are independent of the dark photon masses, and found to be 75\% and 40\% for the 2015--2016 and 2017--2018 data-taking periods, respectively.
These values are estimated considering events with \darky decays into muon pairs with $|\eta|<1.0$ and transverse decay length (\lxy) less than 6~m.
The tri-muon trigger efficiency varies between 10\% and 13\%, depending on the dark photon mass.
 
Unprescaled \MET triggers with the lowest possible threshold were
used to search for displaced \LJs decaying into electrons or light hadrons where triggering on the \LJ itself is difficult,
and to enhance signal acceptance for the displaced \LJs decaying into muons.
Initially, a threshold of 70~\gev\ was used in 2015, but it was raised multiple times afterwards to account for the increasing effects of multiple $pp$ interactions in the same and neighbouring bunch crossings. By the end of the 2017–2018 data-taking period, the threshold was raised to 110~\gev.
 
During $pp$ collisions at the LHC, two counter-rotating proton beams are circulated, consisting of bunches of protons. However, not all bunch slots are necessarily filled with protons after LHC injection, with the number of unfilled bunches dependent on the LHC filling scheme~\cite{Evans:2008zzb}. An empty bunch crossing occurs when neither beam contains protons. During empty bunch crossings, a data sample enriched in cosmic-ray muon background is collected (the ``cosmic data sample''). Additionally, the empty bunches are required to be separated from filled bunches by at least five bunches on each side.
A beam-induced background (BIB)~\cite{DAPR-2012-01} enriched data sample (the ``BIB data sample'') is collected during unpaired isolated bunch crossings. Unpaired bunch crossings happen when only one of the two beams is filled with protons. Additionally, unpaired bunches are required to be separated from filled bunches by at least three unfilled bunches on each side.
 
Monte Carlo (MC) simulated event samples are used to model the BSM signals.
The signal process, where dark photons are produced from the decays of a 125~\gev\ Higgs boson, is simulated
at leading-order (LO) accuracy in $\alpha_\mathrm{S}$ with \MGNLO[2.2.3]\cite{Alwall:2014hca} for the matrix element (ME) calculation,
interfaced to \Pythia[8.186]~\cite{Sjostrand:2007gs} for parton showering (PS) and hadronisation.
The \NNPDF[2.3lo]\cite{Ball:2012cx} parton distribution function (PDF) set is utilised.
The SM cross-section for the VBF Higgs production of $3.78\pm0.08$ pb~\cite{Cepeda:2019klc}
corresponding to $m_{H}=$125.09 \gev, is assumed.
Effects of higher-order QCD corrections on the \pT\ of the Higgs boson are evaluated
using a reweighting procedure~\cite{ATL-PHYS-PUB-2021-020}.
The impact of these effects on the signal selection efficiency
is found to be less than the MC statistical accuracy, and not accounted for in the analysis.
The mean proper decay length $c\tau$ of the \darky is considered a free parameter.
In the simulated samples, \ctau was chosen to ensure that, accounting for the boost of the \darky,
a significant fraction of the decays occur inside the sensitive ATLAS detector volume
(up to the second precision layer in the MS at 7~m in radius and 13~m along the $z$-axis from the centre of the detector).
The mass of \fd was chosen to be much smaller than the Higgs boson mass and far from the kinematic threshold at $m_{\mathrm{HLSP}} + m_{\gamma_{\mathrm{d}}} = m_{f_{\mathrm{d}}}$.
The values of $m_{\mathrm{HLSP}}$ and $m_{f_{\mathrm{d}}}$  have a negligible impact on the analysis results
provided that $m_{\mathrm{HLSP}}$ is much smaller than $m_{\mathrm{H}}$.
The mass of the \darky is chosen to be less than twice the values of $m_{\mathrm{HLSP}}$ to ensure that
it will decay exclusively into SM leptons and light quarks, with the branching fractions depending on its mass.
For example, with a mass of 0.4~\gev, the expected dark photon decay branching ratios are
45\%~\ee, 45\%~$\mu^+\mu^-$, and 10\%~$q\bar{q}$~\cite{Falkowski:2010cm}.
 
The largest backgrounds come from multijet, $\textit{W}+\mathrm{jets}$, $\textit{Z}+\mathrm{jets}$, $\ttbar$, and single-top-quark productions.
Data-driven techniques are used to estimate these backgrounds, with MC simulated samples used
assisting in validation, uncertainty evaluation, and the training of dedicated multivariate classifiers.
The multijet events were simulated using \PYTHIA[8.210]~\cite{Sjostrand:2014zea}
with the same set of tuned parameter values (tune) and PDF used for the signal samples.
The \textit{W}+jets and \textit{Z}+jets samples were simulated using \Sherpa[2.2.1]~\cite{Gleisberg:2008ta}
with the \NNPDF[3.0nnlo]~\cite{Ball:2014uwa} PDF set.
The single-top and \ttbar productions were simulated using \POWHEGBOX[v2]~\cite{Frixione:2007vw} and \PYTHIA[8.230] with the A14 tune~\cite{ATL-PHYS-PUB-2014-021} for PS and hadronisation, and the \NNPDF[2.3lo] set of PDFs.
 
Finally, MC samples of $J/\psi \to \mu\mu$ events were
used to estimate systematic uncertainties for muon trigger and reconstruction efficiencies. These samples were simulated using \Pythia{8} +
\PHOTOSpp{}~\cite{Golonka:2005pn} with the A14 tune~\cite{ATL-PHYS-PUB-2014-021} for PS and hadronisation, and the CTEQ6L1~\cite{Gao:2013xoa,Pumplin:2002vw} PDF set.
 
All MC events went through a full simulation of the ATLAS detector geometry and response~\cite{SOFT-2010-01}
using the \GEANT~\cite{Agostinelli:2002hh} toolkit, taking into account of pile-up and detector response to
interactions occurring in neighbouring bunch crossings from the one producing the hard interaction.
To model the effects of pile-up, simulated inclusive $pp$ events (\PYTHIA[8.210] with the A3 tune~\cite{ATL-PHYS-PUB-2016-017}
and the~\NNPDF[2.3lo] set of PDFs),
were overlaid on each generated hard-scatter event and reweighted to match the conditions of the 2015--2018 data taking.
Reconstructed events are reweighted to reproduce the measured distributions of the number of simultaneous interactions in different data-taking periods. Variations in the trigger threshold as a function of the data taking period are also included in simulation.
Corrections were applied to simulated events to ensure agreement with the object, trigger and identification efficiencies determined from control samples in data.


\FloatBarrier

\section{Event reconstruction}
\label{sec:dpjreco}
The analyses presented in this paper use the same reconstruction and identification
methods for the physics objects as in the \LJ searches in the ggF and \textit{WH} production channels~\cite{EXOT-2019-05}.
 
Candidate events are required to have at least one vertex~\cite{ATL-PHYS-PUB-2015-026} reconstructed from
at least two tracks with \pT\ larger than 500~\MeV\ that are consistent with originating from the beam collision region in the $x$--$y$ plane.
The vertex with the highest sum of squared transverse momenta of associated tracks is considered as the primary vertex of the event.
 
Jets are reconstructed by combining three-dimensional energy clusters measured by the calorimeter~\cite{ATL-PHYS-PUB-2015-036} using the anti-$k_t$ jet clustering algorithm~\cite{Cacciari:2008gp,Fastjet} with a radius parameter of $R=0.4$.
The standard jet calibration is applied using corrections derived from MC simulation and in situ measurements~\cite{PERF-2016-04}.
In this analysis, jets must have $\pT>\SI{20}{\gev}$ and $|\eta|<4.9$,
and to meet fundamental quality criteria designed to reject detector noise and non-collision backgrounds~\cite{ATLAS-CONF-2015-029}.
Jets containing $b$-hadrons, also known as $b$-jets, are identified through a multivariate discriminant that uses
track properties~\cite{ATL-PHYS-PUB-2017-013, FTAG-2018-01},
with a working point that has 70\% efficiency for $b$-jets from top quark decays.
The corresponding rejection factor of light-quark and gluon jets, defined as the inverse of efficiency,
is approximately 300; the corresponding factors for jets containing $c$-hadrons and hadronically decaying $\tau$-leptons are around 38 and 8, respectively.
 
Electron candidates are reconstructed by combining isolated energy deposits in the electromagnetic calorimeter with ID tracks.
To be considered, electrons must satisfy the following criteria: $|\eta|<2.47$, a transverse momentum $\pT>\SI{20}{\gev}$, and satisfy the `Tight' requirement, as defined in Ref.~\cite{EGAM-2018-01}.
Muon candidates are reconstructed in the range of $|\eta|<2.5$ by matching tracks in the MS with those in the ID\@.
To be considered as muons, they must meet the following criteria: a transverse momentum $\pT>\SI{20}{\gev}$ and
satisfy the `Medium' identification requirements as defined in Ref.~\cite{MUON-2018-03}.
For both the electron and muon candidates, the matched tracks are required to have a significance of the transverse impact parameter, $\vert d_0\vert/\sigma(d_0) < 5 (3)$, for electrons (muons).\footnote{The transverse impact parameter, $d_0$, is defined as the distance of closest approach of a track to the beam-line, measured in the transverse plane with an error $\sigma(d_0)$. The longitudinal impact parameter, $z_0$, corresponds to the $z$-coordinate distance between the point along the track at which the transverse impact parameter is defined and the primary vertex.}
The longitudinal impact parameter $z_0$, must satisfy $\vert z_0 \sin\theta\vert < 0.5$~mm, where $\theta$ corresponds to the polar angle of the track relative to the beam-line.
Isolation criteria, based on the scalar sum of \pt\ of ID tracks, are further applied to both the electrons and muons as in Ref.~\cite{EXOT-2019-05}.
To resolve ambiguities that can arise from the independent reconstruction of electron, muon and jet candidates in the detector,
the same overlap removal procedures as in Ref.~\cite{EXOT-2019-05} are applied.
 
The missing transverse momentum vector, \ptmiss, with magnitude, \MET, is calculated as the negative vector sum of the transverse momenta of all identified electrons, muons, and jets, along with an additional component called the `soft term'. The soft term is constructed using all tracks originating from the primary vertex that are not matched to any identified lepton or jet. This approach ensures that \MET is adjusted to achieve the most accurate calibration of leptons and jets, while simultaneously suppressing contributions from pile-up interactions by excluding them from the soft term~\cite{ATL-PHYS-PUB-2015-027,ATLAS-CONF-2018-023}.
 
\subsection{Muonic dark-photon jets}
\label{sec:muonconst}
 
A dark photon decaying into muons outside the ID is expected to generate two or more collimated standalone MS tracks, commonly referred to as a
`muonic dark-photon jet' \mLJ.
Standalone MS tracks~\cite{MUON-2018-03} are reconstructed in the region of MS coverage where the pseudorapidity is $|\eta| \geq 0.1$.
These tracks are formed by identifying at least two matched segments in the MS and are then fit with a primary vertex constraint. Although displaced, the muons are expected to point to the PV due to the large boost. Candidates with an absolute value of the pseudorapidity ranging from $1.0$ to $1.1$ are excluded to avoid the transition region of the MS between the barrel and endcap. Additionally, only standalone MS tracks falling in the pseudorapidity interval $|\eta|<2.5$, which corresponds to the ID coverage, are selected to allow the computation of an isolation variable based on ID tracks.
Standalone MS tracks are required not to match any prompt muon candidate, to discard muons originating from the primary interaction vertex.
 
The \mLJs are reconstructed by a Cambridge--Aachen clustering algorithm~\cite{Dokshitzer:1997in}, combining all the selected standalone MS tracks that lie inside a fixed-size cone in the ($\eta$, $\phi$) space. The reconstruction process begins with the highest-\pt standalone MS track and proceeds by searching for additional standalone MS tracks inside a cone of radius \delr~=~0.4 around the initial track's momentum vector. If a second standalone MS track is found inside this cone, the axis of the cone is adjusted to the vector sum of the momenta of the two tracks. This process is repeated iteratively until no further tracks are found inside the cone.
The \mLJs are required to contain a minimum of two MS tracks and are discarded if a jet is found within a distance of \delr~=~0.4. This last requirement ensures that \mLJs remain distinct from other types of dark-photon jets.
 
Cosmic-ray muons that cross the detector in time coincidence with a $pp$ interaction constitute the main source of background in the \mLJ.
A dense neural network (DNN), referred to as the cosmic-ray tagger, is used to discriminate signal \mLJs~and reject the \mLJ candidates that originate from the cosmic-ray background. The DNN is implemented using Keras with the Tensorflow backend~\cite{tensorflow2015-whitepaper}, with its training setup
and performance detailed in Ref.~\cite{EXOT-2019-05}.
The selection was optimised to obtain a background rejection of 90\% while retaining a high signal efficiency. Signal \mLJs~are selected with an efficiency greater than 95\% for transverse decay lengths \lxy up to 5~m and for \darky~transverse momentum larger than 20~\GeV.

\subsection{Calorimeter dark-photon jets}
\label{sec:hadconst}
 
A dark-photon decaying into electron or quark pairs in the hadronic calorimeters results in energy deposits that are reconstructed as a single jet with a low electromagnetic fraction (EMF), defined as the ratio of the energy deposited in the EM calorimeter to the total jet energy. Jets with an EMF below $0.4$ are referred to as calorimeter dark-photon-jet candidates \hLJs.
Low EMF jets are reconstructed and calibrated using the same algorithms mentioned earlier for standard jets. However, they are only considered if they have $\pT>\SI{20}{\gev}$ and lie in the range of $|\eta|<2.5$.
Candidates in the transition region between the barrel calorimeters and the endcap cryostat are removed by requiring the fraction of energy in the Tile Gap scintillators to be less than 10\% of the total jet energy.
 
To maintain high efficiency for the targeted signals, low EMF jets are required to satisfy quality criteria that are
less stringent than those in the standard jet selection.
To reject fake \hLJs\ arising from prompt jets, low EMF jets with more than 40\% of matched tracks compatible with originating from the primary vertex, as defined by the jet vertex tagger (JVT)~\cite{PERF-2014-03} are removed.
Further \hLJs cleaning is applied as detailed in Ref.~\cite{EXOT-2019-05} to reject background from noise bursts~\cite{LARG-2013-01}.
These cleaning requirements result in the rejection of approximately 0.8\% of low EMF jets in the signal samples.

The time matched to a \hLJ, denoted by $t_{\mathrm{\hLJ}}$,
is measured as the energy-weighted average of the timing for each calorimeter cell matched to the jet,
corrected by the corresponding time-of-flight from the interaction point. It is used to reject cosmic-ray muons and BIB by requiring
events from $pp$ collisions to have a value of $|t_{\mathrm{\hLJ}}|$ that is in a window of 4 ns around zero.
 
To reduce fake \hLJs arising from prompt jets, a dedicated discriminator (\hLJtagger), based on a convolutional neural network implemented using Keras with the Tensorflow backend, is used to assign a score to each \hLJ in the event. The training of the neural network and its setup  are discussed in Ref.~\cite{EXOT-2019-05}. The \hLJtagger inputs are three-dimensional representations of energy deposits matched to the jet. The energy deposits are defined by collections of calorimeter cell clusters used in jet reconstruction~\cite{PERF-2014-07}. Each collection has $(\eta,\phi)$ coordinates and holds information about the total amount of energy deposited in each of the calorimeter samplings.
A \hLJ is accepted if the \hLJtagger output is larger than 0.5, indicating an enhanced probability of a displaced signature.
This corresponds to a selection efficiency of more than 70\% for \darky\ with transverse decay lengths \lxy in the range of
2--3.5 m and \pT~larger than 20~\gev, with a background rejection of 94\%~\cite{EXOT-2019-05}.
 
Muons arising from BIB can deposit energy in the HCAL by radiative losses, which can be reconstructed as \hLJs owing to the resulting low EMF. To reduce the residual contamination from misidentified \hLJs from BIB, a dedicated per-jet tagger (\hLJBIBtagger) as detailed in Ref.~\cite{EXOT-2019-05} is used.
The selection is optimised to obtain a signal efficiency of greater than 80\% with a corresponding BIB rejection of 68\%.


\FloatBarrier

\section{Event selection and background estimate}
\label{sec:analysis}
 
Long-lived dark photon candidate events are selected by requiring at least one \LJ satisfying the selection criteria
described in Section~\ref{sec:dpjreco}.
If more than one \LJ is reconstructed, only the one with the highest transverse momentum, called the `leading \LJ',
is considered and used to classify the event into two exclusive categories, \hLJs and \mLJs, based on its type.
In the \mLJ channel, events are triggered by the logical OR of the dedicated muon triggers (tri-muon and narrow-scan)
and the \etmiss trigger as discussed in Section~\ref{sec:samples},
to target both the displaced muonic signature and the large \etmiss\ in the FRVZ signal models.
In the \hLJ channel, events are triggered by the \etmiss trigger.
 
To reject background from SM processes and to select
events consistent with the VBF production, events are required to have:
\begin{itemize}
\item at least two jets with $\pt \geq 30$~\GeV;
\item the invariant mass of the two leading jets, \mjj$\geq 1000$~\GeV;
\item the pseudo-rapidity gap between the two leading jets, $|\detajj|>3$;
\item the azimuthal angular difference between the two leading jets, $|\dphijj|<2.5$;
\item exactly zero leptons;
\item exactly zero b-tagged jets;
\item $\met\geq 100$~\GeV.
\end{itemize}
 
The centrality in pseudo-rapidity of the \LJ between the two VBF jets, \cLJ, is used to enhance the
analysis sensitivity in the \mLJ channel. It is defined as:
\begin{equation}
\cLJ = \exp\Big( - \frac{4}{(\eta_\mathrm{j1} - \eta_\mathrm{j2})^2}\big(\eta_\mathrm{DPJ}-\frac{\eta_\mathrm{j1} +\eta_\mathrm{j2}}{2}\big)^2\Big). \notag
\label{eq:centraliyu}
\end{equation}
In the \mLJ signal region (SR), \SRm, the value of \cLJ is required to be greater than 0.7,
to capture the VBF topology where the \LJ is centred between the two VBF jets.
To further reduce the multijet background events in the \hLJ channel,
the minimum azimuthal distance between the \ptmiss and the $\vec{p}_{\mathrm{T}}$ of each of the four
leading jets in the event, $\dphimin$, is required to be greater than 0.4
to reject events with mismeasured jet energies leading to fake \met\ in the event.
 
To improve the analysis sensitivity, the \hLJ final state is further split into two exclusive SRs, \SRhl and \SRhh,
where $\met$ is in the range of [100, 225]~\GeV and $\etmiss$ is $>225$~\GeV, respectively.
Table~\ref{tab:SR} summarises the selections applied in all three SRs.
Further selections, as shown in Table~\ref{tab:SR}, are applied to the leading \LJ to improve the search sensitivity based on the
following variables:
\begin{itemize}
\item The \mLJ charge is the sum of the charges of muon-tracks matched to a \mLJ is sensitive to the charge of the initiating state. \mLJs originating from neutral particle decays are expected to have zero sum of charge;
\item The \dpjiso\ is the scalar sum of the transverse momenta of all tracks within a $\Delta R=0.5$ cone around the direction of the \LJ\ momentum vector. Displaced \LJs are expected to have very little nearby track activity in the ID.
 
\end{itemize}

\begin{table}[!h]
\caption{\label{tab:SR}
Definition of the signal regions in the \mLJ and \hLJ channels.
Selections on the \LJ are applied only to the leading one.
The \hLJ final state is further split into two signal regions, \SRhl, and \SRhh, with
\etmiss values of $\in[100, 225]\, \GeV$ and $>225\, \GeV$ respectively.
Dashes indicate cases where a requirement is not applied.}
\def\arraystretch{1.2}
\begin{center}
\begin{tabular}{l c c}
\toprule
Requirement / Region              &  \SRm  &  \SRhlhh   \\
\noalign{\smallskip}\hline\noalign{\smallskip}
Number of \LJs                    & $\geq 1$ & $\geq 1$ \\
Leading DPJ type                  & \mLJ       & \hLJ \\
\noalign{\smallskip}\hline\noalign{\smallskip}
\multirow{3}{*}{Trigger}          &  \etmiss   &  \multirow{3}{*}{\etmiss}  \\
& Tri-muon MS-only &  \\
& Muon narrow-scan  &  \\
\noalign{\smallskip}\hline\noalign{\smallskip}
$\pt(\mathrm{jet}) \;[\GeV]$      &  $>30$ & $>30$ \\ 
\njet                             & $\geq 2$ & $\geq 2$ \\ 
$\mjj \;[\GeV]$                   & $\geq 1000$ & $\geq 1000$ \\ 
$|\detajj|$                       & $>3$      & $>3$ \\ 
$|\dphijj|$                       & $<2.5$    & $<2.5$ \\
\midrule
$N_{\ell}$     									  & 0 & 0 \\
\nbjet                            & 0 & 0 \\ 
\midrule
\cLJ                              & $>0.7$   & -    \\
$\dphimin$                   			&  -       & $>0.4$ \\
\multirow{2}{*}{$\met \;[\GeV]$}  & \multirow{2}{*}{$>100$}   & \SRhl: $[100, 225]$ \\
&          & \SRhh: $>225$ \\
|\mLJ charge|                       & 0        & -  \\
\hLJ QCD tagger                       & -        &  $>0.9$  \\
\dpjiso~[GeV]                     & $<2$     &  $<2$ \\
\hline
\end{tabular}
\end{center}
\end{table}


\begin{table}[!h]
\caption{\label{tab:crsr} Definition of the control regions used in the background estimation.
All CRs requirements are the same as for the respective SR, with the exception of the selections reported in this table.
End points within the ranges are included (excluded) when denoted by square (round) brackets.
}
\def\arraystretch{1.2}
\begin{center}
\begin{tabular}{l r r r}
\noalign{\smallskip}\hline\noalign{\smallskip}
Requirement / Region      & \CRBm         & \CRCm         & \CRDm     \\
\noalign{\smallskip}\hline\noalign{\smallskip}
|\mLJ charge|             & $[1, 5)$     & $[1, 5)$      &   0 \\
\dpjiso[GeV]             & $[0,2.0)$       & $[2.0, 20)$   & $[2.0, 20)$        \\
\noalign{\smallskip}\hline\noalign{\smallskip}
Requirement / Region      & \CRBhlhh       & \CRChlhh         & \CRDhlhh     \\
\noalign{\smallskip}\hline\noalign{\smallskip}
\hLJ QCD tagger score     & $[0.9,1]$        & $[0.8, 0.9)$  & $[0.8, 0.9)$ \\
\dpjiso~[GeV]             & $[2.0 , 20)$  & $[2.0 , 20)$  & $[0,2.0)$        \\
\hline
\end{tabular}
\end{center}
\end{table}


The main sources of background left after the SR selections are punch-through jets
from rare multijet events for \SRm, and multijet and electroweak $W$ and $Z$ production
for \SRhl\ and \SRhh.
 
Non-collision backgrounds, including cosmic-ray muon and BIB, are found to be negligible.
The cosmic-ray muon background contribution is estimated from events collected with the analysis triggers in the cosmic data sample.
By requiring at least one reconstructed  \mLJ, the contribution from cosmic-ray muon is found to be subdominant.
After all selection, it is found to be negligible in all signal regions.
Events with BIB energy deposits are very likely to have a \hLJ which can contribute to the background.
The possible contribution from misidentified \hLJs was studied using the BIB-enriched data sample.
The BIB contamination is found to be negligible after the timing and BIB tagger requirements.
Hence this background contribution in the SR is neglected.
 
As nearly all background in the SR arises from a prompt jet misidentified as a displaced jet,
a common data-driven `ABCD' method is used to estimate their contributions.
This method relies on the assumption that the distribution of background events can be factorised
in two dimensions using two uncorrelated variables.
The background events are then subdivided into four regions: A, B, C, and D.
In the ABCD convention, region A is the SR, regions B, C, D are also referred to
as control regions CRB, CRC, and CRD\@.
Without any signal contamination in regions B, C, and D,
the number of background events in the SR can be estimated
as $N_\mathrm{A} = N_\mathrm{D} \times N_\mathrm{B}/ N_\mathrm{C}$, where
$N_\mathrm{B,C,D}$ are the number of observed events in data in the B, C, and D regions.
Any possible signal leakage outside the SR region is accounted for by using a modified ABCD method that simultaneously
fits the signal and background events in all regions.
The two dimensional distributions used in the estimate of the backgrounds are formed by \{\mLJ charge, \dpjiso\} in the \mLJ channel,
and \{\hLJtagger\ discriminant, \dpjiso\} in the \hLJ channel.
Definitions of the control regions CRB, CRC, and CRD are shown in Table~\ref{tab:crsr}. All CRs requirements are the same as for the respective SR, with the exception of the selections reported in this table.
 
The background estimate procedure was validated by applying the
ABCD method in validation regions (VRs), where signal contributions are small.
A first set of VRs are obtained by combining control regions as defined in Table~\ref{tab:crsr},
such as a combination of CRB and CRC, or a combination of CRC and CRD\@.
To improve the statistical uncertainty, the selection requirements on \etmiss\ in the \mLJ\ VRs is relaxed to be greater than 20~\GeV,
and the VRs in the \hLJ\ channel are defined within a common \etmiss\ range of $\etmiss>100$~\GeV.
A second set of the VRs are defined by inverting the selections on $\dphijj$ or $\dphimin$ in the \mLJ or \hLJ channel, respectively.
To improve the statistical uncertainty in these VRs, the \cLJ\ selection requirement is removed in the \mLJ\ channel, and
the \hLJ\ VRs are defined within a common \etmiss\ range of $\etmiss>100$~\GeV.
 
For each VR, alternative (ABCD)$'$ test regions are defined, and
region boundaries are varied in discrete steps to assess the dependence on the variables used in the ABCD planes.
The Pearson linear correlation coefficient between the two variables defining the ABCD plane in all
VRs was found to be less than 3\% and the signal leakage  was found to be less than 1\% of the total signal in the SR selection for all signal scenarios considered in the analysis.
In the \mLJ\ VRs, \dpjiso\ boundaries are tested in the range of 0.5--4.5~\gev\ or 2.5--7.0~\gev.
In the \hLJ\  VRs, \dpjiso\ boundaries are tested in the range of 0.5--4.5~\gev\ or
2.5--5.5~\gev, and boundaries on the \hLJtagger\ discriminant boundaries are tested
in the range of 0.84--0.98 or 0.82--0.88.
In all the VRs, the observed yields were found to agree with
the estimated background predictions within about one standard deviation considering
only the statistical uncertainties in the VRs.
The relative statistical uncertainties of the predicted backgrounds in the VRs
are about 25\% and 15\% in the \mLJ\ and \hLJ\ channels, respectively, when using
the same boundary definitions as in the SRs.

Figure~\ref{fig:ABCD_unblinded_mudpj} shows the distribution of data events in the ABCD plane for the \mLJ channel, together with the expected distribution for a benchmark FRVZ model assuming a 125~\gev\ Higgs boson and a decay branching fraction to the dark sector of 10\%.
The acceptance times efficiency for the FRVZ signal processes after applying all SR selection criteria is 0.31\% in the case of a \darky mass of 400~\mev\ and a generated \ctau of 50~mm.
 
\begin{figure}[h]
\centering
\subfloat[]{\includegraphics[width=0.4\textwidth]{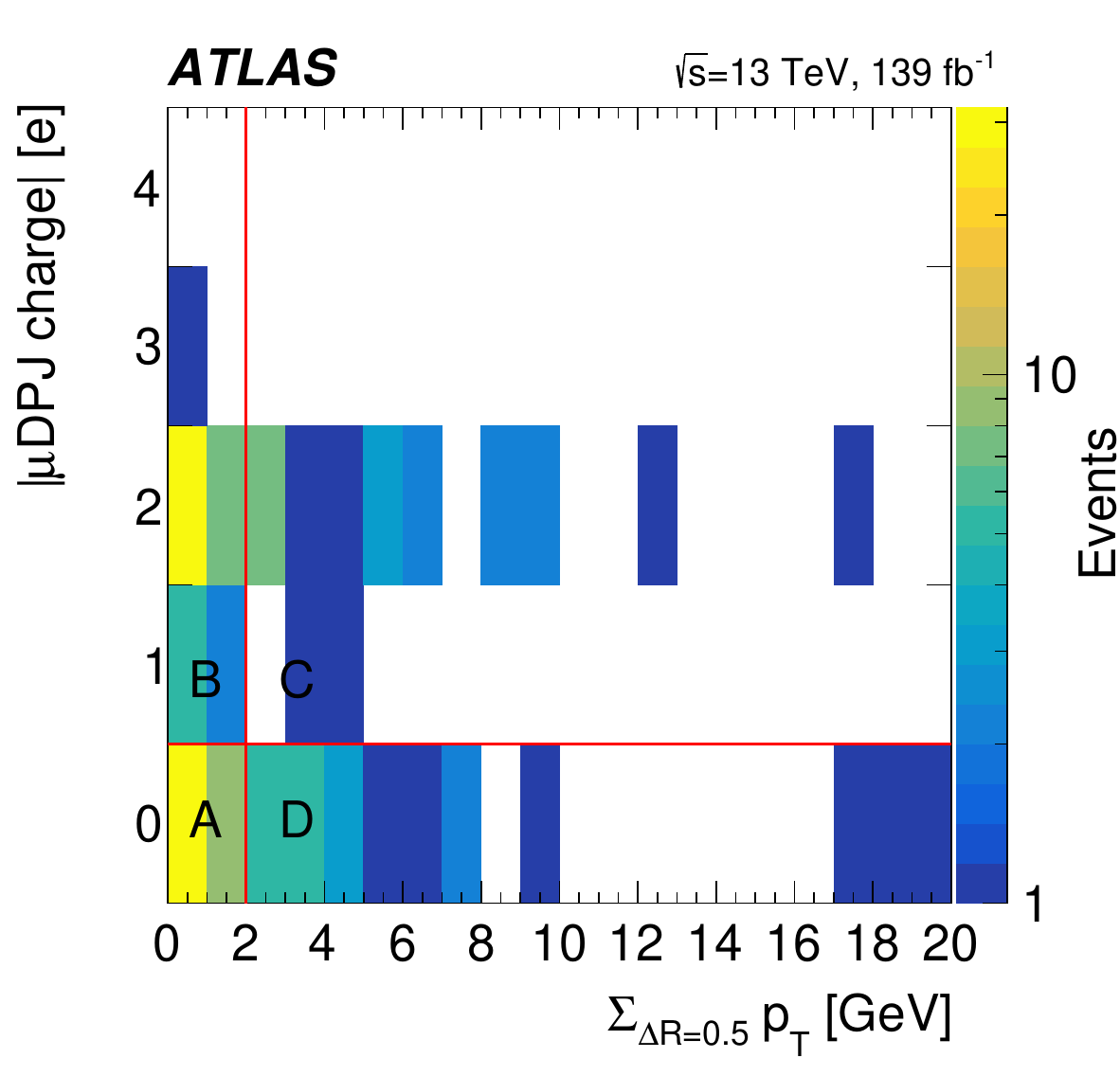}}
\subfloat[]{\includegraphics[width=0.4\textwidth]{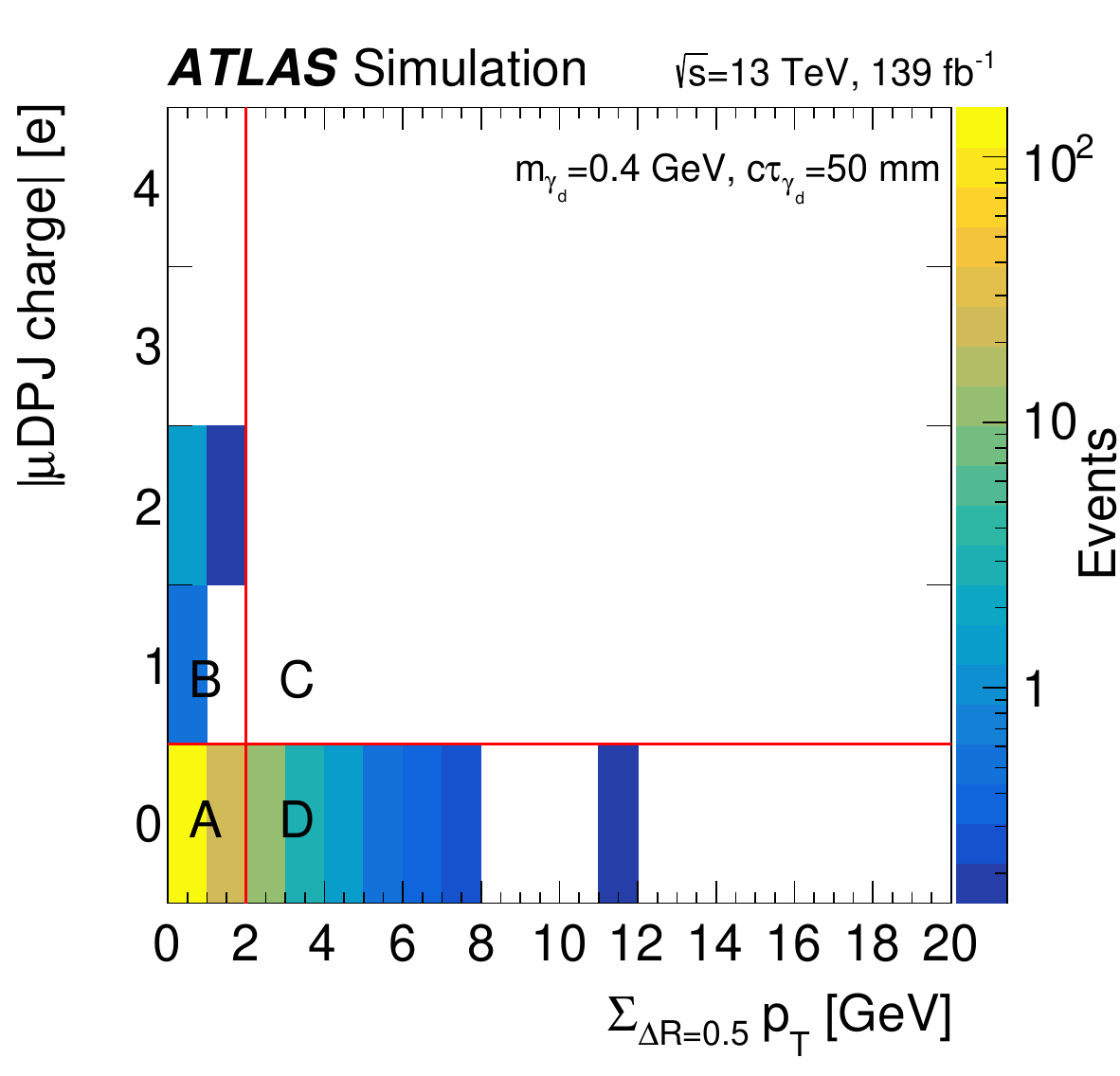}}
\caption{
The per-event \mLJ\ charge vs \dpjiso\ distributions in the ABCD planes defined for the \SRm. Figure (a) shows events in data, while Figure (b) shows simulated FRVZ events, assuming a 125~\gev\ Higgs boson with a decay branching fraction of 10\% to the dark sector and \darky with mass of 400~\mev\ and \ctau=50~mm.}
\label{fig:ABCD_unblinded_mudpj}
\end{figure}

Figure~\ref{fig:ABCD_unblinded_calodpj} shows the distribution of data events in the ABCD plane for the \hLJ channels, together with the expected distribution for a benchmark FRVZ model assuming a 125~\GeV\ Higgs boson and a decay branching fraction to the dark sector of 10\%.
The acceptance times efficiency for the FRVZ signal processes after applying SR selection criteria
is 0.093\% or 0.032\% for \SRhl\ or \SRhh,
in the case of a \darky mass of 100~\MeV\ and a generated \ctau of 15~mm.

\begin{figure}[h]
\centering
\subfloat[]{\includegraphics[width=0.4\textwidth]{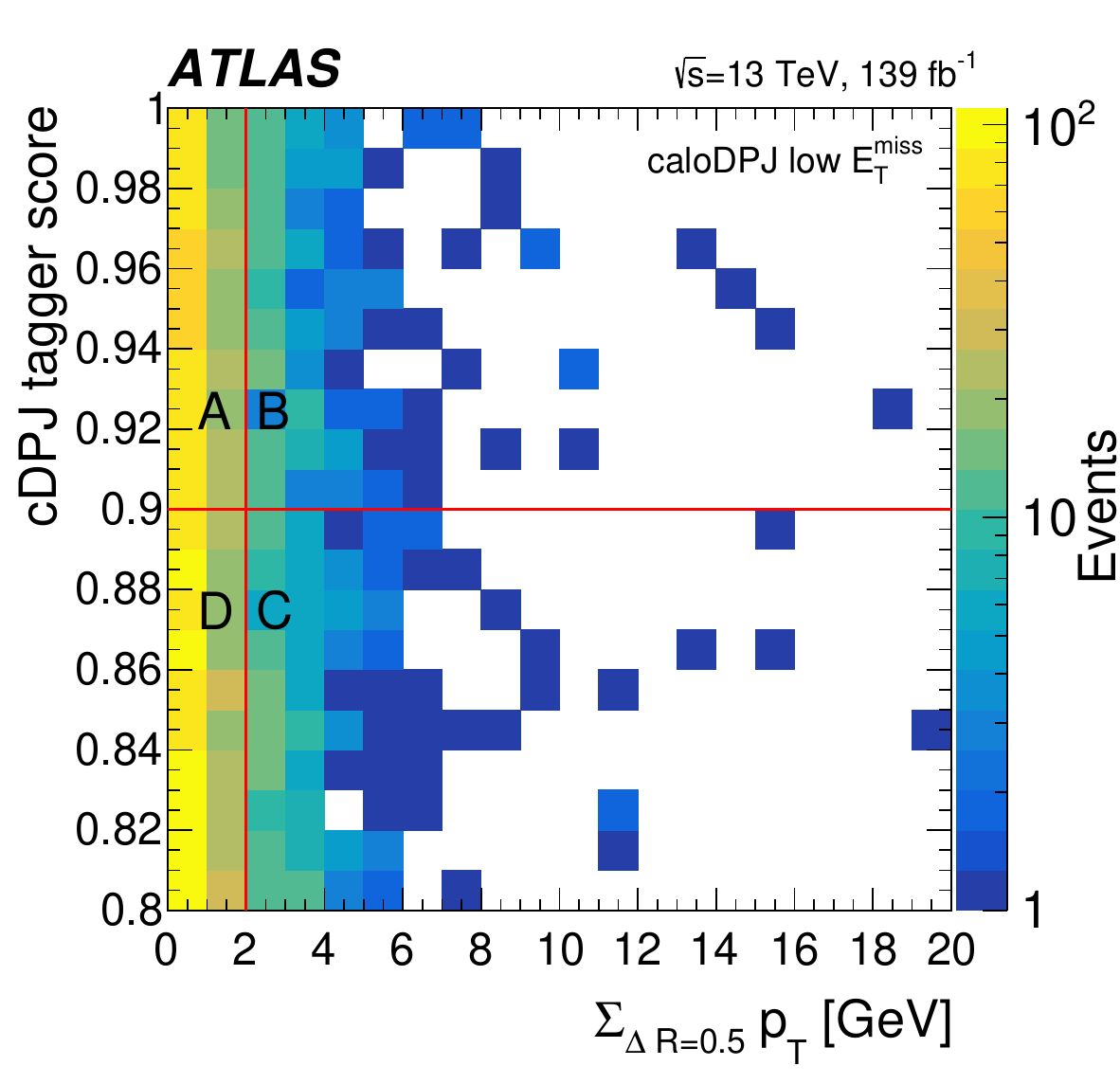}}
\subfloat[]{\includegraphics[width=0.4\textwidth]{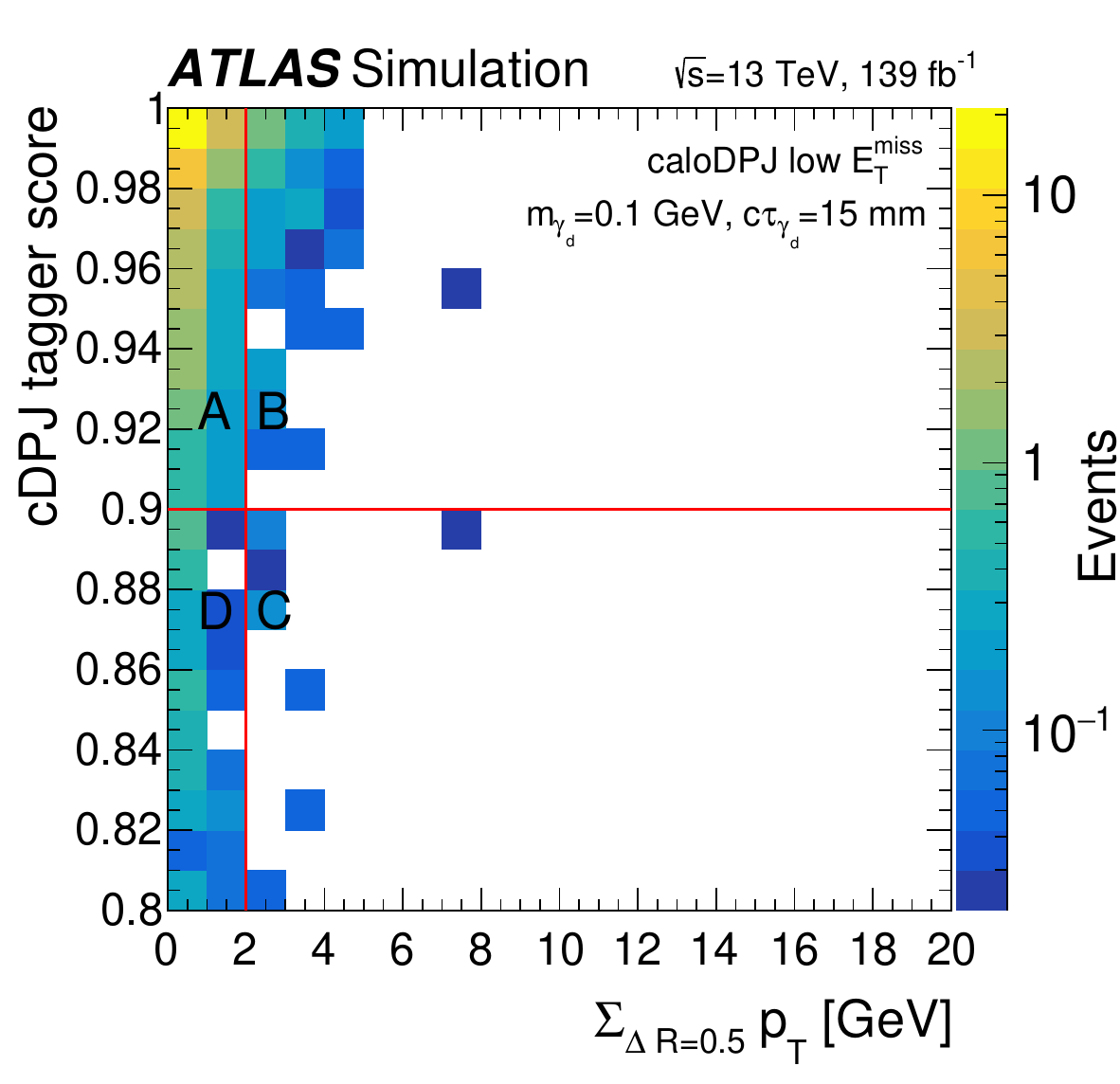}}\\
\subfloat[]{\includegraphics[width=0.4\textwidth]{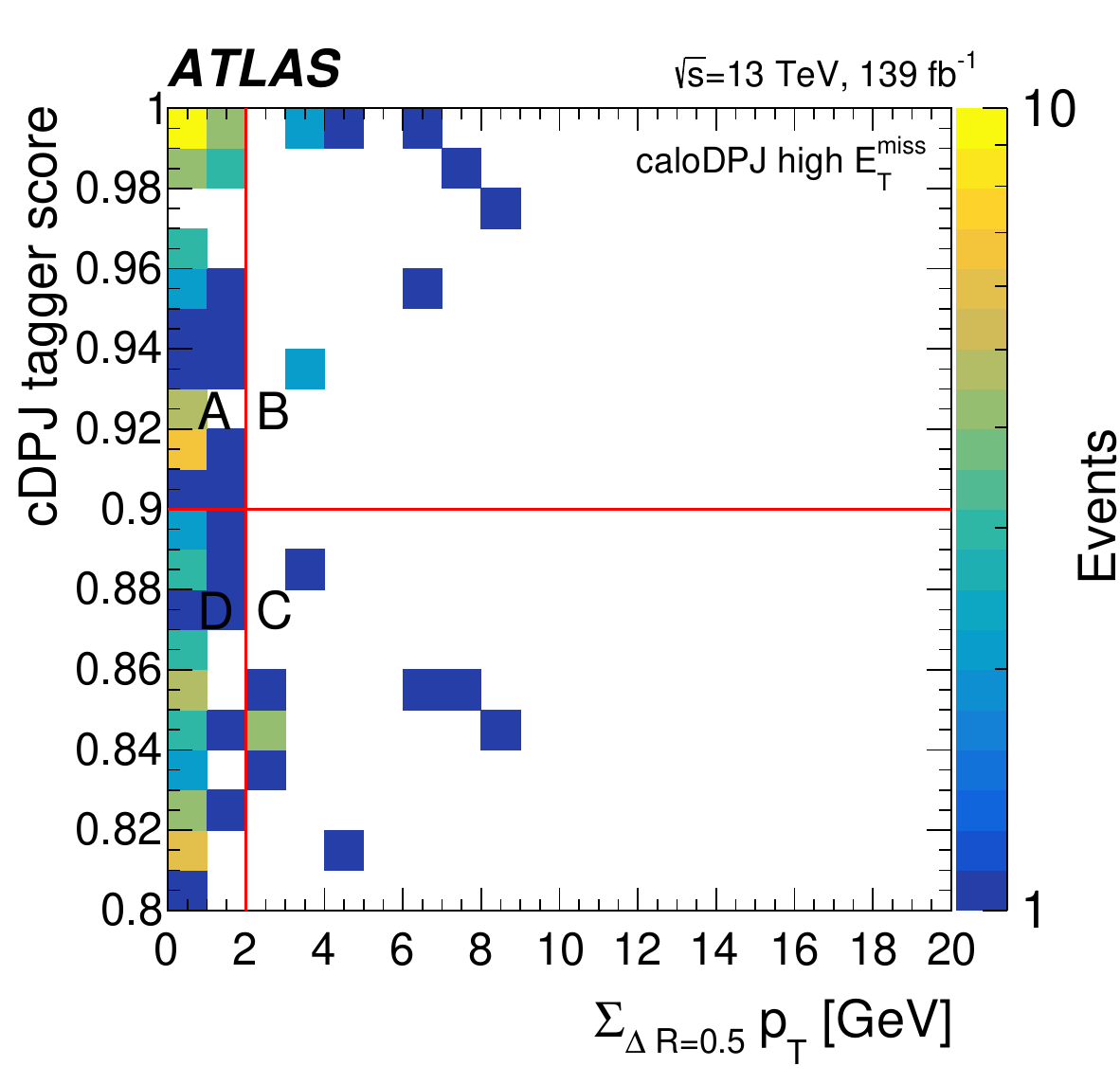}}
\subfloat[]{\includegraphics[width=0.4\textwidth]{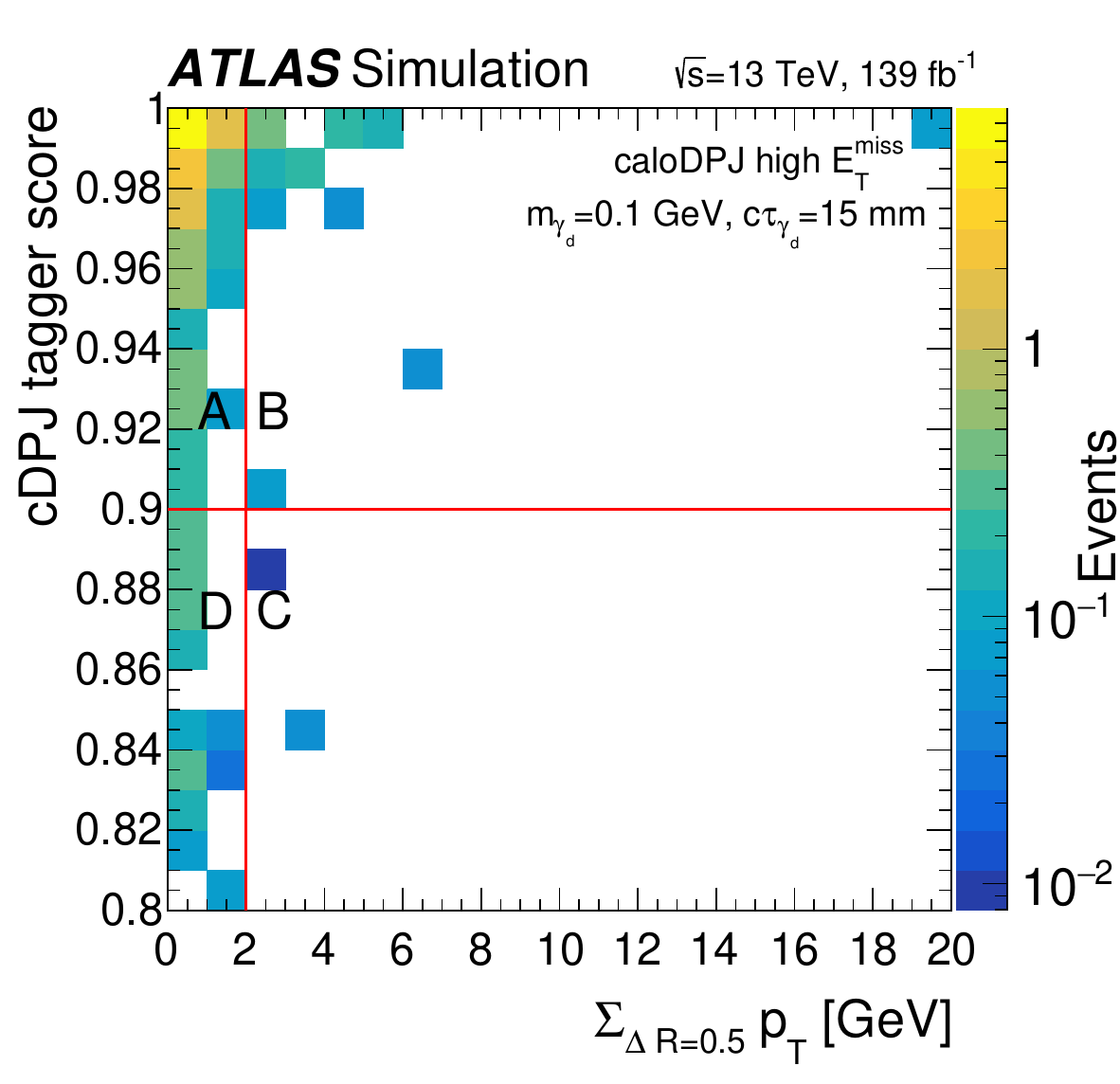}}
\caption{
The per-event \hLJtagger\ vs \dpjiso\ distributions in the ABCD planes defined for the (a,b) \SRhl\ and (b,c) \SRhh\ channels.
Figures (a,c) show events in data, while Figures (b,d) show simulated FRVZ events,
assuming a 125~\GeV\ Higgs boson with a decay branching fraction of 10\% to the dark sector and \darky with mass of 100~\mev\ and \ctau=15~mm.}
\label{fig:ABCD_unblinded_calodpj}
\end{figure}

\FloatBarrier


\FloatBarrier

\section{Systematic uncertainties}
\label{sec:uncertainties}
Potential sources of experimental uncertainties
are considered for the simulated signal yields to account for potential data and MC differences in the event reconstruction and selections.
For the background estimations, the statistical uncertainties in the observed yields in regions CRB, CRC, and CRD are directly
propagated to the SR from the ABCD method.
 
Experimental uncertainties for jets include jet energy scale (JES) and jet energy resolution (JER) uncertainties
from the standard calibration scheme~\cite{PERF-2016-04}, which amount to less than 7\% of the expected signal yields.
An additional JES uncertainty is applied to consider a possible dependence on the low-EM-fraction selection. It is estimated following the same procedure as used in the 2015--2016 dark-photon jet search~\cite{EXOT-2017-28} and is up to 13\% for the \hLJ channels.
 
In the \mLJ channel, a systematic uncertainty in the single-\darky reconstruction efficiency is evaluated using a
tag-and-probe method applied to $J/\psi \to \mu\mu$ events in data and simulation, taking into account the
leading effect from the opening angle \delr between the two muons.
The $J/\psi$ reconstruction efficiency is evaluated in both the data and simulation as a function of
the opening angle \delr between the two muons, and the corresponding difference between data and simulation
in the \delr~$<0.06$ region, where the \LJ samples are concentrated, is found to be up to 9.6\% and taken as the uncertainty.
 
The systematic uncertainty in the \etmiss trigger efficiency is evaluated by propagating the statistical error of the trigger scale-factors introduced in Section~\ref{sec:samples}.
It is estimated to be less than 9\% and 4\% in \SRhl and \SRm respectively.
 
The uncertainties related to the MC modelling of the taggers are obtained from Ref.~\cite{EXOT-2019-05} and recalculated for all signal samples. In the case of the cosmic-ray tagger, the calculated uncertainty is found to be less than 4\% in \SRm. Conversely, for the \hLJtagger and the \hLJBIBtagger, the estimated uncertainties are found to be less than 6\% and 12\%, respectively, in both the \SRhl and \SRhh.
 
A pile-up modelling uncertainty is assigned to account for the difference between the predicted and measured inelastic cross-sections~\cite{STDM-2015-05}, which is evaluated to be less than 5\% in all signal regions.
 
A systematic uncertainty in the procedure used for extrapolating the signal efficiency to different lifetimes was evaluated by comparing
the extrapolated efficiency derived from the main simulated samples with the measured efficiency of samples with alternative lifetime assumptions.
The results showed agreement within statistical uncertainties, except for the \SRhl selection, where the extrapolated efficiency at large lifetime values is lower by 36\% from the nominal value. Consequently, a one-sided systematic
uncertainty of 36\% is included when extrapolating events from the \SRhl region to lifetimes larger than the generated MC lifetime.
 
Finally, an uncertainty in the computed total integrated luminosity used to rescale the expected number of signal events is considered. The uncertainty in the combined 2015--2018 integrated luminosity is 1.7\%~\cite{ATLAS-CONF-2019-021}, obtained using the LUCID-2 detector~\cite{LUCID2} for the primary luminosity measurements.
 
Figure~\ref{fig:sysPlot} summarises the average experimental uncertainties for the bulk of the signal samples,
where absolute deviations are found to be on the order of a few percent up to 6\% for the total uncertainty across different \darky\ masses.
 
\begin{figure}[ht!]
\centering
\includegraphics[width=.95\textwidth]{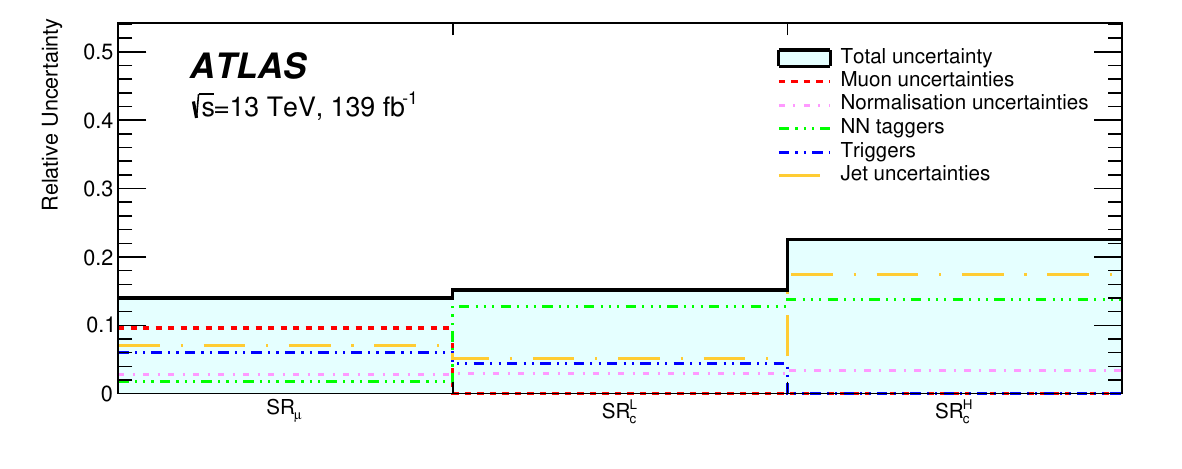}
\caption{ Comparison of the relative uncertainty in the signal yield in each SR, showing the contributions from the different sources of uncertainty summed in quadrature. The quoted values are averaged over different \darky\ masses. The `Muon uncertainties' category contains all muon-related systematic uncertainties and is dominated by the uncertainty in the single-\darky\ reconstruction efficiency. The `NN taggers' category contains the three taggers adopted in the analysis and is dominated by the BIB tagger uncertainties for the calorimetric channels and by the cosmic-ray tagger in the muonic channel. The `Triggers' category contains all trigger systematic uncertainties. The `Jet uncertainties' category contains the JES, JER and low-EM-fraction JES systematic uncertainties. Some sets of systematic uncertainties apply to only a subset of the SRs. The systematic uncertainty in the lifetime extrapolation is not shown.
}
\label{fig:sysPlot}
\end{figure}


\FloatBarrier

\section{Results and interpretations}
\label{sec:result}
A combined maximum-likelihood fit to the yields in the four  (i.e.\ ABCD) regions in data is used for
data-driven background estimations in each SR and for model interpretations.
The likelihood function is constructed as a product of Poisson functions, one for each of the SR, CRB, CRC, and CRD regions,
describing signal and background expectations. The ABCD ansatz is introduced as nuisance parameters in the background component of the expected yield in each region.
The likelihood-based ABCD fit is robust against control regions with only a few events and considers possible signal contamination in the control regions.
Higgs-boson contribution via the ggF mechanism or in association with a vector boson (\textit{WH} or \textit{ZH}) are at the percent level, and thus neglected.
The effects of systematic uncertainties described in Section~\ref{sec:uncertainties} are treated in the fit by nuisance parameters with Gaussian constraints.
These parameters are assumed to be uncorrelated across regions,
where an alternative correlation model with full correlation across regions is studied and found to have a negligible impact on the results.
The mean value of the Gaussian probability distribution function is constrained by the nominal value of the parameter and the variance is defined by the 68\% confidence interval of the systematic uncertainty associated with the parameter.
 
Table~\ref{tab:DataAbcd} presents the observed and expected numbers of events in all SRs.
The reported yields are extrapolated by the fit assuming no signal, and with unblinded data in all ABCD regions in the fit.
When comparing the results with a likelihood fit using blinded data in the SR, the background yields are found to be consistent within percent level.
No significant excess above expected background is observed.
 
\begin{table}[h]
\caption{Observed and expected yields in the ABCD regions. The total uncertainty in the background expectation is computed by the ABCD background-only fit to unblinded data. }
\label{tab:DataAbcd}
\def\arraystretch{1.2}
\centering
\begin{tabular}{cccccc}
\hline
Selection   & CRB    & CRC    & CRD      & SR expected     & SR observed \\  \hline
\hline
\SRm        &  44      &   22     &     21     & ~~~$42 \pm 14$  & ~~~41       \\
\hline
\SRhl       &   224     &    256    &    1123      & ~~$983 \pm 95$  & ~~923       \\
\SRhh       &  9      &    11    &     35     & ~~~$29 \pm 14$  & ~~46       \\
\hline
\end{tabular}
\end{table}
 
The results are then used to set upper limits on the branching fraction \bHtwogd
as a function of the \darky mean proper decay length \ctau,
assuming the SM VBF production cross section of a 125~\GeV\ Higgs boson,
with $X$ assumed to be undetected.
The upper limit on the signal strength, relative to the predicted signal yields from simulation, is obtained with the CL$_\text{s}$ method~\cite{Read:2002hq} with the asymptotic calculator~\cite{Cowan:2010js}.
The validity of the asymptotic approximation is checked against a full calculation using pseudo-experiments,
and the CL$_\text{s}$ values of the two methods typically agree within 5\%.
 
Figure~\ref{fig:limits_vbf} shows a summary of the observed limits on the branching ratio \bHtwogd\ assuming the production of a 125~\gev\ Higgs boson.
The sensitivity of each SR and the combination are reported in separate subfigures for the VBF production.
In the region where \darky\ masses are less than twice the muon mass,
dark photons can decay into electrons and light quarks within the muon spectrometer volume and are
reconstructed as \mLJ.
Sensitivities in the \mLJ\ channel are typically much better than the \hLJ\ channels due to the larger signal efficiency, except
in the region where dark photon masses are below twice the muon mass.
The high \etmiss\ region in the \hLJ\ channel mainly contributes in scenarios involving dark photons with masses larger than 10~\GeV and large \ctau values, where one of the dark photons may decay outside ATLAS.
 
The results in the VBF analysis are then combined with the SRs in the ggF and \textit{WH} analyses~\cite{EXOT-2019-05},
constructed to be mutually exclusive by different selections on the \mjj\ and the number of prompt leptons.
The combined fit considers a product of the likelihood functions of the individual search channels with independent nuisance parameters,
but with a common signal normalisation. Systematic uncertainties in the SRs are assumed to be fully uncorrelated.
In each production channel, contributions from other Higgs boson productions are relatively small,
and not considered as signal for simplicity.
Figure~\ref{fig:limits_combi} shows the expected and observed limits
on \bHtwogd\ for a \darky\ mass of 0.1 and 10~\GeV, as functions
of the dark photon mean proper decay length \ctau, combining all production channels.
Observed limits in different SRs used in the combination are also presented for comparison.
The VBF channel contributes significantly to the sensitivity for the dark photons
with short and long decay lengths.
Branching fractions larger than 5\% can be excluded at
95\% CL if the dark photons have a mean proper decay length \ctau\ between 1 mm and 5 mm and a mass below twice the muon mass.
For the scenarios where dark photon masses are between 0.4~\gev\ and 6~\gev, branching fractions larger than 1\% can be excluded at
95\% CL if the dark photons have a mean proper decay length \ctau\ between 1 mm and 267 mm.
For the dark photons with masses of 10~\gev, branching fractions larger than 10\% are excluded at
95\% CL if the dark photons have a mean proper decay length \ctau\ between 173 mm and 1296 mm.

Upper limits at 90\% CL\footnote{The limits are quoted at 90\% CL in this interpretation of the results to ease the comparison with other experiments that conventionally use this CL threshold.} are also set, in the context of the FRVZ model in terms of kinetic mixing parameter $\epsilon$ and \darky mass and presented in Figure~\ref{fig:DMexclusionNew} for different \bHtwogd, ranging from 0.1\%~to~10\%.
The limits are interpolated between different masses by branching fraction variations~\cite{Batell:2009yf} as a function of the \darky mass, corrected by a linear interpolation of the signal efficiency between adjacent available MC signal samples.
For a \darky mass below twice the muon mass, no coverage from \mLJ signal regions is expected, motivating a significant drop in sensitivity.

\begin{figure}[h!]
\centering
\subfloat[]{\includegraphics[width=.49\textwidth]{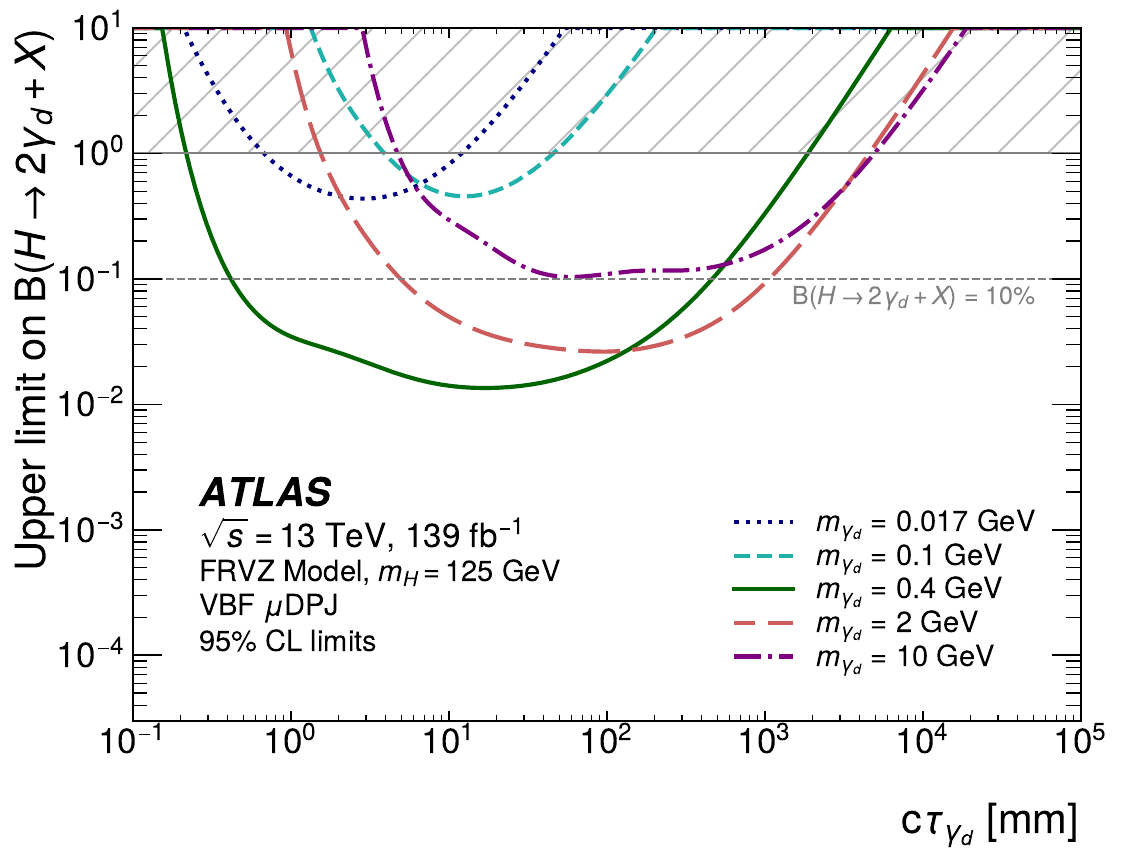}}
\subfloat[]{\includegraphics[width=.49\textwidth]{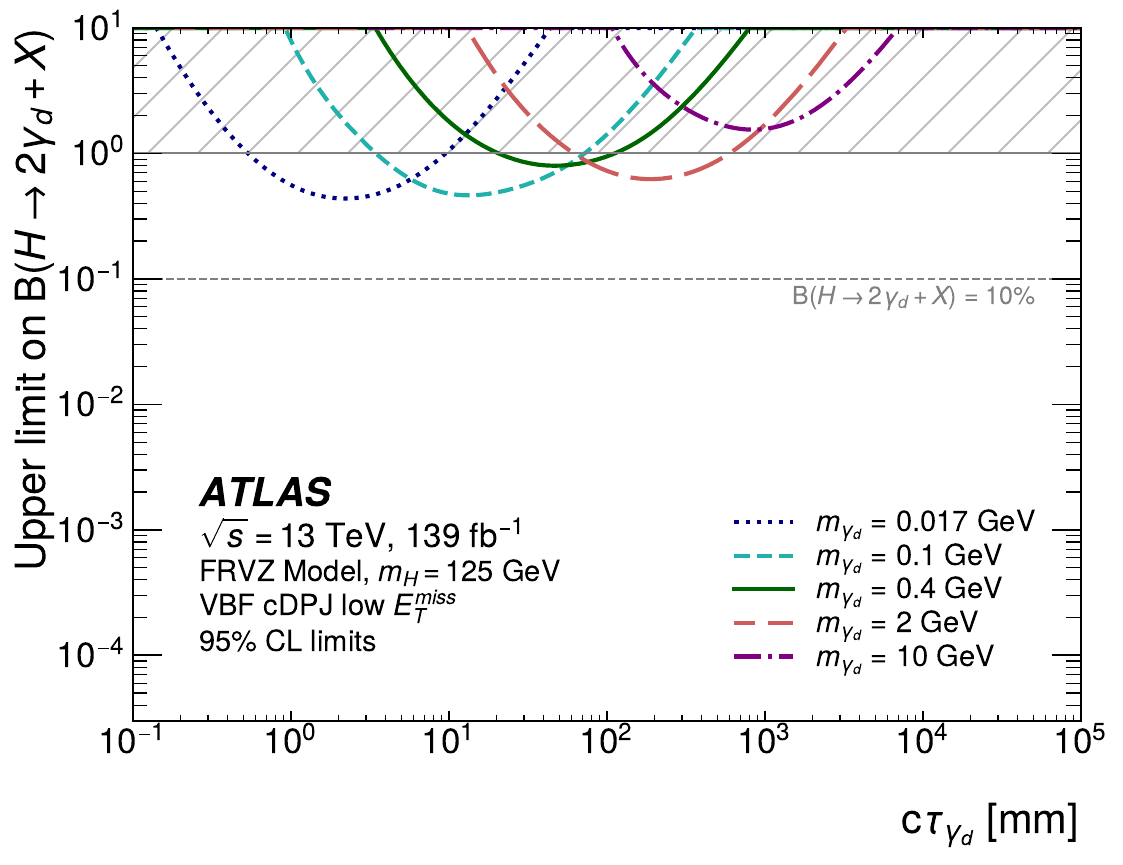}}\\
\subfloat[]{\includegraphics[width=.49\textwidth]{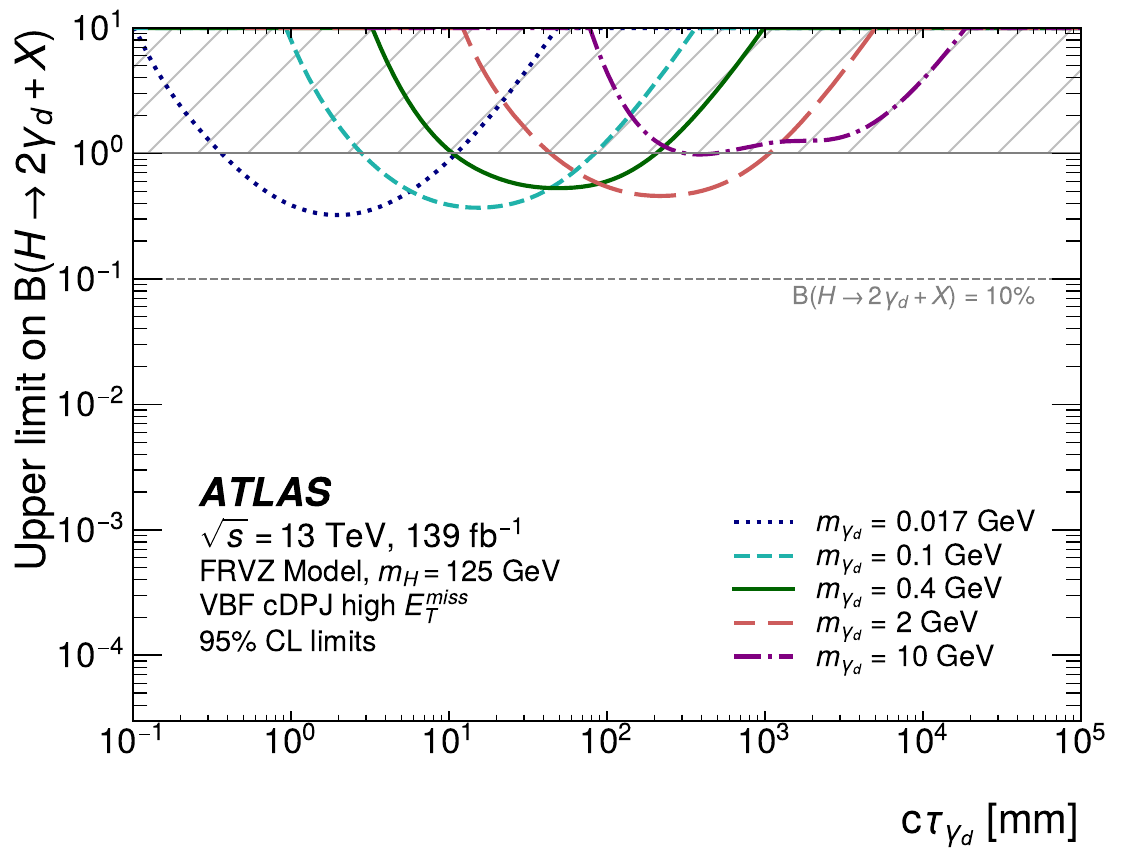}}
\subfloat[]{\includegraphics[width=.49\textwidth]{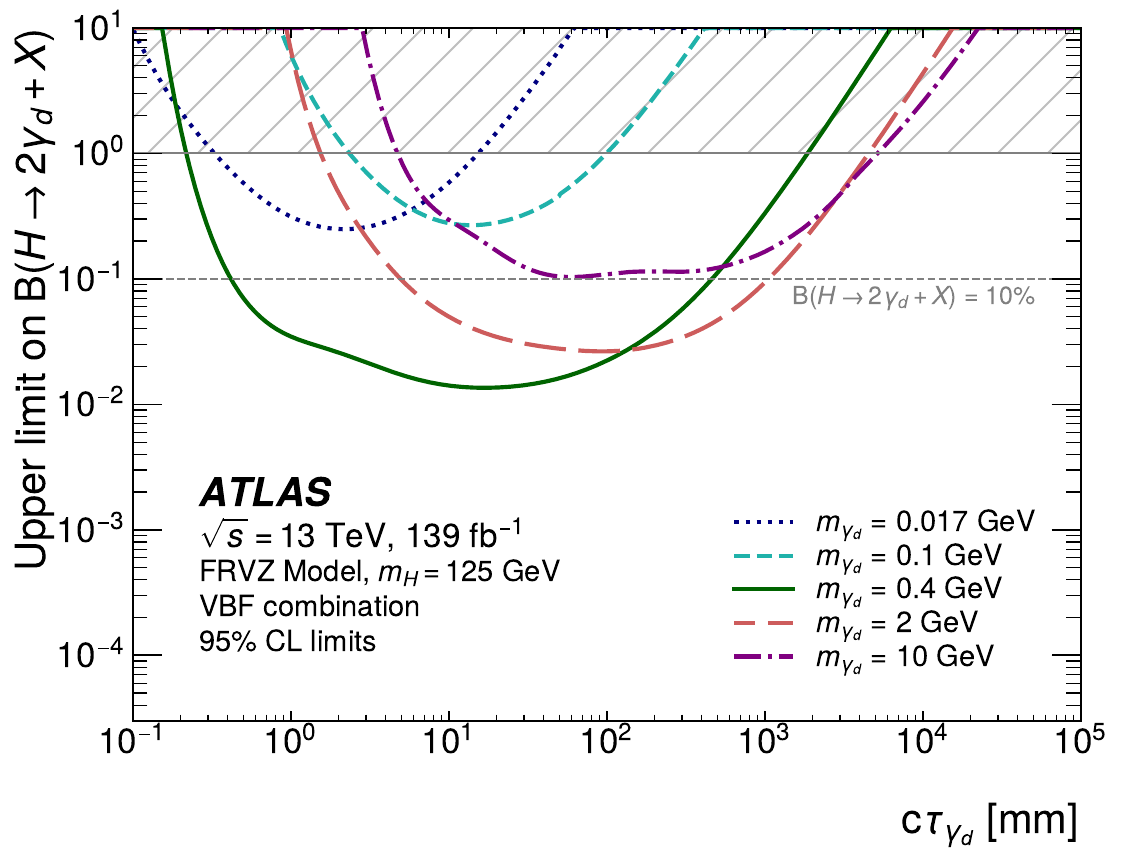}}
\caption{Upper limits at 95$\%$ CL on the branching fraction as a function of the dark-photon mean proper decay length \ctau for the \Htwogd process, assuming a 125~\gev\ Higgs boson and a FRVZ signal model.
The limits are shown separately for the (a)~\SRm, (b)~\SRhl, (c)~\SRhh search channels, and (d) with all channels combined. The hatched band denotes the region in which the branching ratio is larger than unity.}
\label{fig:limits_vbf}
\end{figure}

\begin{figure}[h!]
\centering
\subfloat[]{\includegraphics[width=.49\textwidth]{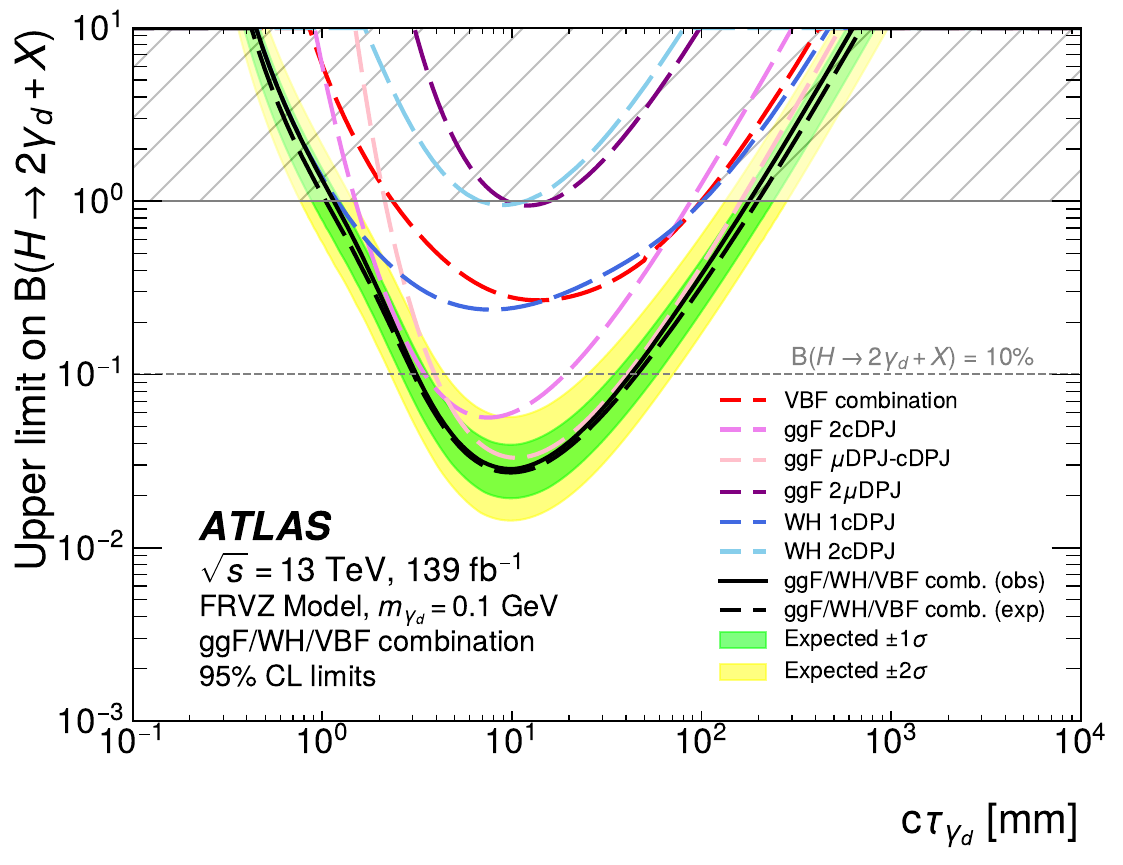}}
\subfloat[]{\includegraphics[width=.49\textwidth]{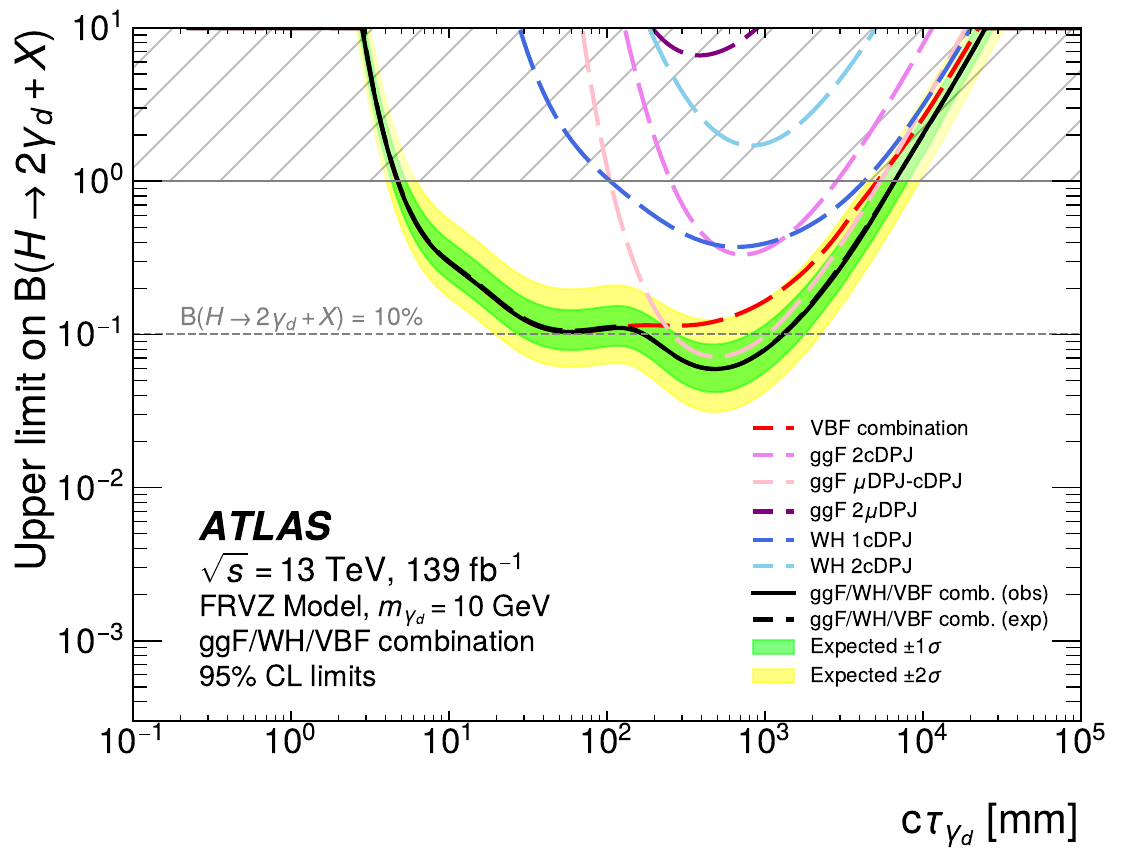}}
\caption{
Upper limits at 95$\%$ CL on the branching fraction as a function of the dark-photon mean proper decay length \ctau for the \Htwogd process, assuming a 125~\gev\ Higgs boson and a FRVZ signal model.
The limits are shown separately for each channel and for the statistical combination (solid black) for a \darky\ mass of (a) 100~\mev\ and (b) 10~\gev.
The hatched band denotes the region in which the branching ratio is larger than unity.         }
\label{fig:limits_combi}
\end{figure}

\begin{figure}[ht]
\centering
\includegraphics[width=.75\textwidth]{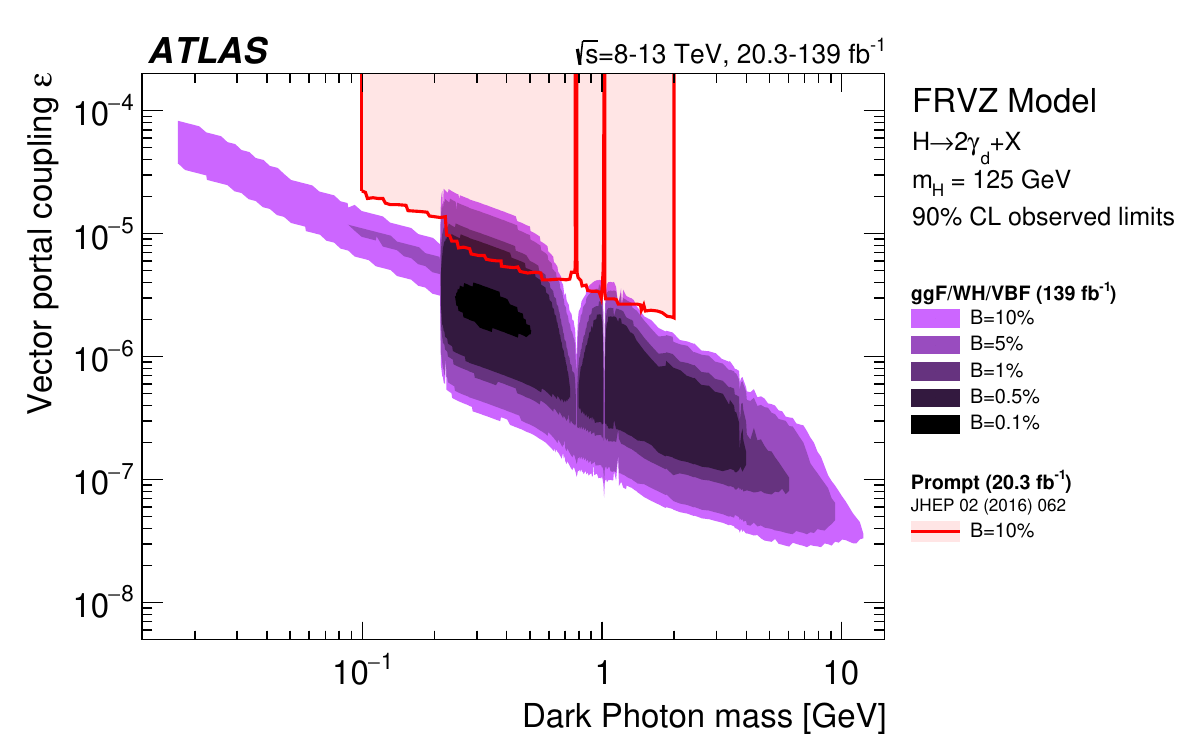}
\caption{90$\%$ CL exclusion contours of the \bHtwogd\ as a function of the \darky\ mass and the kinetic mixing parameter $\epsilon$.
These limits are obtained assuming branching fractions between 0.1\% and 10\% (shown as distinct shades of purple within the contour regions) for Higgs boson decays resulting in dark photons.
The figure also shows the region excluded by the previous ATLAS search for prompt~\cite{EXOT-2014-09} (solid red line) decays of dark photons.
}
\label{fig:DMexclusionNew}
\end{figure}


\FloatBarrier

\section{Conclusion}
\label{sec:conclusion}
A search for long-lived dark photons decaying into collimated pairs of fermions
is performed using 139~\ifb of data collected from $pp$ collisions at $\sqrt{s}=13$\ \tev\ at the ATLAS detector.
The analysis considers the FRVZ signal models where dark photons are produced from Higgs boson decays.
The VBF topology, which is characterised by two energetic jets that have a large rapidity separation in addition
to large values of \mjj\ and \etmiss\, is studied for the first time  to target the VBF Higgs boson production mode.
The reduction of background in the VBF channel makes it feasible to
require only one long-lived dark photon candidate in the events, sensitive
to the dark photons with shorter or longer decay lengths where one of the dark photons
may decay promptly or outside ATLAS and escape the \LJ\ reconstruction.
The data are found to be consistent with the background prediction.
Upper limits on \bHtwogd\ as a function of dark photon mass and mean proper decay length \ctau are reported, assuming
the SM cross-section for a 125 \gev\ Higgs boson.
 
This search is also combined with previous ATLAS searches obtained in the gluon--gluon fusion and \textit{WH} production modes~\cite{EXOT-2019-05}.
Branching fractions above 10\% can be excluded at 95\% CL for \Htwogd\ decays
for dark photons with mean proper decay length between 173 and 1296~mm and mass of 10~\gev,
compared with the range of [263--1030] mm in the previous publication~\cite{EXOT-2019-05}.
The improvement is entirely due to the addition of the VBF channel and the statistical combination.


\FloatBarrier
 
\section*{Acknowledgements}


We thank CERN for the very successful operation of the LHC and its injectors, as well as the support staff at
CERN and at our institutions worldwide without whom ATLAS could not be operated efficiently.
 
The crucial computing support from all WLCG partners is acknowledged gratefully, in particular from CERN, the ATLAS Tier-1 facilities at TRIUMF/SFU (Canada), NDGF (Denmark, Norway, Sweden), CC-IN2P3 (France), KIT/GridKA (Germany), INFN-CNAF (Italy), NL-T1 (Netherlands), PIC (Spain), RAL (UK) and BNL (USA), the Tier-2 facilities worldwide and large non-WLCG resource providers. Major contributors of computing resources are listed in Ref.~\cite{ATL-SOFT-PUB-2023-001}.
 
We gratefully acknowledge the support of ANPCyT, Argentina; YerPhI, Armenia; ARC, Australia; BMWFW and FWF, Austria; ANAS, Azerbaijan; CNPq and FAPESP, Brazil; NSERC, NRC and CFI, Canada; CERN; ANID, Chile; CAS, MOST and NSFC, China; Minciencias, Colombia; MEYS CR, Czech Republic; DNRF and DNSRC, Denmark; IN2P3-CNRS and CEA-DRF/IRFU, France; SRNSFG, Georgia; BMBF, HGF and MPG, Germany; GSRI, Greece; RGC and Hong Kong SAR, China; ISF and Benoziyo Center, Israel; INFN, Italy; MEXT and JSPS, Japan; CNRST, Morocco; NWO, Netherlands; RCN, Norway; MEiN, Poland; FCT, Portugal; MNE/IFA, Romania; MESTD, Serbia; MSSR, Slovakia; ARRS and MIZ\v{S}, Slovenia; DSI/NRF, South Africa; MICINN, Spain; SRC and Wallenberg Foundation, Sweden; SERI, SNSF and Cantons of Bern and Geneva, Switzerland; MOST, Taipei; TENMAK, T\"urkiye; STFC, United Kingdom; DOE and NSF, United States of America.
 
Individual groups and members have received support from BCKDF, CANARIE, CRC and DRAC, Canada; CERN-CZ, PRIMUS 21/SCI/017 and UNCE SCI/013, Czech Republic; COST, ERC, ERDF, Horizon 2020, ICSC-NextGenerationEU and Marie Sk{\l}odowska-Curie Actions, European Union; Investissements d'Avenir Labex, Investissements d'Avenir Idex and ANR, France; DFG and AvH Foundation, Germany; Herakleitos, Thales and Aristeia programmes co-financed by EU-ESF and the Greek NSRF, Greece; BSF-NSF and MINERVA, Israel; Norwegian Financial Mechanism 2014-2021, Norway; NCN and NAWA, Poland; La Caixa Banking Foundation, CERCA Programme Generalitat de Catalunya and PROMETEO and GenT Programmes Generalitat Valenciana, Spain; G\"{o}ran Gustafssons Stiftelse, Sweden; The Royal Society and Leverhulme Trust, United Kingdom.
 
In addition, individual members wish to acknowledge support from Chile: Agencia Nacional de Investigaci\'on y Desarrollo (FONDECYT 1190886, FONDECYT 1210400, FONDECYT 1230987); China: National Natural Science Foundation of China (NSFC - 12175119, NSFC 12275265); European Union: European Research Council (ERC - 948254), Horizon 2020 Framework Programme (MUCCA - CHIST-ERA-19-XAI-00), Italian Center for High Performance Computing, Big Data and Quantum Computing (ICSC, NextGenerationEU), Marie Sklodowska-Curie Actions (EU H2020 MSC IF GRANT NO 101033496); France: Agence Nationale de la Recherche (ANR-20-CE31-0013, ANR-21-CE31-0013, ANR-21-CE31-0022), Investissements d'Avenir Idex (ANR-11-LABX-0012), Investissements d'Avenir Labex (ANR-11-LABX-0012); Germany: Baden-Württemberg Stiftung (BW Stiftung-Postdoc Eliteprogramme), Deutsche Forschungsgemeinschaft (DFG - CR 312/5-1); Italy: Istituto Nazionale di Fisica Nucleare (FELLINI G.A. n. 754496, ICSC, NextGenerationEU); Japan: Japan Society for the Promotion of Science (JSPS KAKENHI 22H01227, JSPS KAKENHI JP21H05085, JSPS KAKENHI JP22H04944); Netherlands: Netherlands Organisation for Scientific Research (NWO Veni 2020 - VI.Veni.202.179); Norway: Research Council of Norway (RCN-314472); Poland: Polish National Agency for Academic Exchange (PPN/PPO/2020/1/00002/U/00001), Polish National Science Centre (NCN 2021/42/E/ST2/00350, NCN OPUS nr 2022/47/B/ST2/03059, NCN UMO-2019/34/E/ST2/00393, UMO-2020/37/B/ST2/01043, UMO-2021/40/C/ST2/00187); Slovenia: Slovenian Research Agency (ARIS grant J1-3010); Spain: BBVA Foundation (LEO22-1-603), Generalitat Valenciana (Artemisa, FEDER, IDIFEDER/2018/048), La Caixa Banking Foundation (LCF/BQ/PI20/11760025), Ministry of Science and Innovation (RYC2019-028510-I, RYC2020-030254-I), PROMETEO and GenT Programmes Generalitat Valenciana (CIDEGENT/2019/023, CIDEGENT/2019/027); Sweden: Swedish Research Council (VR 2018-00482, VR 2022-03845, VR 2022-04683), Knut and Alice Wallenberg Foundation (KAW 2017.0100, KAW 2018.0157, KAW 2018.0458, KAW 2019.0447); Switzerland: Swiss National Science Foundation (SNSF - PCEFP2\_194658); United Kingdom: Leverhulme Trust (Leverhulme Trust RPG-2020-004); United States of America: Neubauer Family Foundation.


\clearpage
 
\printbibliography
 
\clearpage
 
\begin{flushleft}
\hypersetup{urlcolor=black}
{\Large The ATLAS Collaboration}

\bigskip

\AtlasOrcid[0000-0002-6665-4934]{G.~Aad}$^\textrm{\scriptsize 102}$,
\AtlasOrcid[0000-0002-5888-2734]{B.~Abbott}$^\textrm{\scriptsize 120}$,
\AtlasOrcid[0000-0002-1002-1652]{K.~Abeling}$^\textrm{\scriptsize 55}$,
\AtlasOrcid[0000-0001-5763-2760]{N.J.~Abicht}$^\textrm{\scriptsize 49}$,
\AtlasOrcid[0000-0002-8496-9294]{S.H.~Abidi}$^\textrm{\scriptsize 29}$,
\AtlasOrcid[0000-0002-9987-2292]{A.~Aboulhorma}$^\textrm{\scriptsize 35e}$,
\AtlasOrcid[0000-0001-5329-6640]{H.~Abramowicz}$^\textrm{\scriptsize 151}$,
\AtlasOrcid[0000-0002-1599-2896]{H.~Abreu}$^\textrm{\scriptsize 150}$,
\AtlasOrcid[0000-0003-0403-3697]{Y.~Abulaiti}$^\textrm{\scriptsize 117}$,
\AtlasOrcid[0000-0002-8588-9157]{B.S.~Acharya}$^\textrm{\scriptsize 69a,69b,m}$
% ...
\bigskip
\\

% ...
$^{29}$Physics Department, Brookhaven National Laboratory, Upton NY; United States of America.\\
% ...
$^{35}$$^{(a)}$Facult\'e des Sciences Ain Chock, R\'eseau Universitaire de Physique des Hautes Energies - Universit\'e Hassan II, Casablanca;$^{(b)}$Facult\'{e} des Sciences, Universit\'{e} Ibn-Tofail, K\'{e}nitra;$^{(c)}$Facult\'e des Sciences Semlalia, Universit\'e Cadi Ayyad, LPHEA-Marrakech;$^{(d)}$LPMR, Facult\'e des Sciences, Universit\'e Mohamed Premier, Oujda;$^{(e)}$Facult\'e des sciences, Universit\'e Mohammed V, Rabat;$^{(f)}$Institute of Applied Physics, Mohammed VI Polytechnic University, Ben Guerir; Morocco.\\
% ...
$^{49}$Fakult\"{a}t Physik , Technische Universit{\"a}t Dortmund, Dortmund; Germany.\\
% ...
$^{55}$II. Physikalisches Institut, Georg-August-Universit\"{a}t G\"ottingen, G\"ottingen; Germany.\\
% ...
$^{69}$$^{(a)}$INFN Gruppo Collegato di Udine, Sezione di Trieste, Udine;$^{(b)}$ICTP, Trieste;$^{(c)}$Dipartimento Politecnico di Ingegneria e Architettura, Universit\`a di Udine, Udine; Italy.\\
% ...
$^{102}$CPPM, Aix-Marseille Universit\'e, CNRS/IN2P3, Marseille; France.\\
% ...
$^{117}$Department of Physics, New York University, New York NY; United States of America.\\
% ...
$^{120}$Homer L. Dodge Department of Physics and Astronomy, University of Oklahoma, Norman OK; United States of America.\\
% ...
$^{150}$Department of Physics, Technion, Israel Institute of Technology, Haifa; Israel.\\
$^{151}$Raymond and Beverly Sackler School of Physics and Astronomy, Tel Aviv University, Tel Aviv; Israel.\\
% ...
$^{a}$ Also Affiliated with an institute covered by a cooperation agreement with CERN.\\
$^{b}$ Also at An-Najah National University, Nablus; Palestine.\\
$^{c}$ Also at Borough of Manhattan Community College, City University of New York, New York NY; United States of America.\\
$^{d}$ Also at Center for High Energy Physics, Peking University; China.\\
$^{e}$ Also at Center for Interdisciplinary Research and Innovation (CIRI-AUTH), Thessaloniki; Greece.\\
$^{f}$ Also at Centro Studi e Ricerche Enrico Fermi; Italy.\\
% ...
$^{m}$ Also at Department of Physics, King's College London, London; United Kingdom.\\
% ...

\end{flushleft}


\end{document}